\documentclass[reprint,nofootinbib,superscriptaddress,amsmath,amssymb,aps,prd]{revtex4-2}
\usepackage{amsmath,mathrsfs,amsbsy,color,graphicx,bm,amsthm,amsfonts}
\usepackage{bbm}
\usepackage{times}
\usepackage{units}
\usepackage{dcolumn}
\usepackage{graphicx}
\usepackage{epsfig}
\usepackage{epstopdf}
\usepackage[colorlinks,linkcolor=red,anchorcolor=green,citecolor=blue,CJKbookmarks=True]{hyperref}
\DeclareMathSymbol{\shortminus}{\mathbin}{AMSa}{"39}
\usepackage{mathrsfs}
\usepackage{braket}
\usepackage{amssymb}
\usepackage{txfonts}
\usepackage{float}
\usepackage{enumitem}
\usepackage{multirow}
\usepackage{orcidlink}
\usepackage{upgreek}

\newcommand{\pp}{\partial}

\newcommand{\eq}[1]{Eq. (\ref{#1})}
\newcommand{\meq}[1]{(\ref{#1})}
\DeclareRobustCommand{\scopeedit}[1]{#1}

\begin{document}
\title{Morphology-resolved Bayesian inference of complex non-Kerr black hole shadows}

\author{Wentao Liu\orcidlink{0009-0008-9257-8155}}
\email[]{liuwentao@lzu.edu.cn} 
\affiliation{Lanzhou Center for Theoretical Physics, Key Laboratory of Theoretical Physics of Gansu Province, 
Key Laboratory of Quantum Theory and Applications of MoE,
Gansu Provincial Research Center for Basic Disciplines of Quantum Physics, 
Lanzhou University, Lanzhou 730000, China}
\affiliation{Institute of Theoretical Physics $\&$ Research Center of Gravitation,
Lanzhou University, Lanzhou 730000, China}

\author{Yu-Xiao Liu\orcidlink{0000-0002-4117-4176}}
\email[]{liuyx@lzu.edu.cn (corresponding author)} 
\affiliation{Lanzhou Center for Theoretical Physics, Key Laboratory of Theoretical Physics of Gansu Province, 
Key Laboratory of Quantum Theory and Applications of MoE, Gansu Provincial Research Center for Basic Disciplines of Quantum Physics, 
Lanzhou University, Lanzhou 730000, China}
\affiliation{Institute of Theoretical Physics $\&$ Research Center of Gravitation,
Lanzhou University, Lanzhou 730000, China}

\begin{abstract}

Black hole shadows provide a direct probe of spacetime geometry, yet existing inference frameworks often treat the full parameter space as a single inference domain, which may obscure morphology-dependent response structures.
Here, we show that this uniform-response picture breaks down for an exact Kerr–Newman-like exterior solution in a Lorentz-violating vector-tensor theory.
A nonvanishing radial–angular metric component modifies photon dynamics, producing droplet-like and strongly distorted shadows together with a spin-reversal asymmetry absent in the standard Kerr–Newman spacetime. 
To render these unconventional morphologies statistically tractable, we develop a shadow-area observable directly from ray-traced images, requiring no analytic representation of the shadow boundary, and use it in Bayesian inference with current EHT measurements of M87*.
Within the parameter range explored, the shadow family evolves continuously in appearance but separates into three response sectors with distinct parameter degeneracies and EHT-constrained posterior structures. 
Continuous changes in visual morphology can therefore mask qualitative transitions in parameter response and the distinct inference regimes they produce. 
These results motivate morphology-resolved Bayesian inference beyond a single global fit.

\end{abstract}

\vspace*{0.5cm}
\maketitle

\section{Introduction}

A black hole is analogous to a coin placed beneath a sheet of paper: although it cannot be observed directly, its presence can be inferred from the way it modifies the surrounding pattern. 
In the astrophysical context, the observed intensity distribution depends on the external emission environment, whereas the underlying photon-capture geometry is determined by the spacetime.
Under suitable observational and modeling assumptions, the resulting dark region and its boundary can therefore provide information about the geometry near the black hole. 
Shadow-based tests of gravity can thus be formulated as an inverse problem: a gravitational model predicts a family of shadow boundaries, while observations assign statistical weights to the corresponding parameter configurations. 
By combining theoretical shadow observables with horizon-scale measurements, one can infer the spacetime parameters and constrain the underlying gravitational model.
The Event Horizon Telescope (EHT) has resolved horizon-scale ring-like emission around M87* \cite{EventHorizonTelescope:2019dse,EventHorizonTelescope:2019uob,EventHorizonTelescope:2019jan,EventHorizonTelescope:2019ths,EventHorizonTelescope:2019pgp,EventHorizonTelescope:2019ggy} and Sgr A* \cite{EventHorizonTelescope:2022wkp,EventHorizonTelescope:2022apq,EventHorizonTelescope:2022wok,EventHorizonTelescope:2022xqj}, providing the first observational basis for such shadow-based tests.
These observations have underscored the importance of theoretical and numerical studies of light rings, accretion and emission physics \cite{Cunha:2020azh,Guo:2020qwk,Davelaar:2021eoi,Cunha:2022gde,Uniyal:2025uvc,Cunha:2018acu,Gralla:2019xty,Perlick:2021aok,Konoplya:2021slg,Vagnozzi:2022moj,Chen:2022scf,Ayzenberg:2023hfw,Gao:2023mjb,Guo:2024cts,Zeng:2025olq,Zhang:2025lhm}.
More recently, the multi-epoch EHT observations of M87* from 2017, 2018, and 2021 revealed a persistent ring diameter of $43.9\pm0.6~\mu{\rm as}$ \cite{EventHorizonTelescope:2025vum}. 
The improved angular-size measurement provides a more precise observational input for testing theoretical shadow models and constraining deviations from the Kerr geometry. 
At the same time, the increased observational precision places greater demands on the parameter-inference methodology: uncertainties and degeneracies among multiple spacetime and viewing parameters must be treated simultaneously rather than through isolated parameter scans.

Most existing shadow constraints are obtained by varying one deformation parameter, or a small subset of parameters, while fixing the remaining black-hole and observational quantities \cite{Vagnozzi:2019apd,Bambi:2019tjh,Long:2019nox,Guo:2018kis,Afrin:2021imp,Meng:2022kjs,Xavier:2023exm,Liu:2024lbi,Liu:2024soc,Liu:2024iec,Tsukamoto:2024gkz,Liu:2025lwj,Ali:2025beh,Ou:2026ohc}. 
Such scans are useful for identifying qualitative trends and deriving model-dependent bounds, but they become increasingly difficult to interpret in a multidimensional parameter space. 
Because different parameters often affect the same shadow observables in similar ways, they can generate intricate degeneracies that cannot be captured by a sequence of one- or two-parameter comparisons.
Bayesian methods have increasingly been applied to horizon-scale imaging data to account for observational uncertainties and parameter degeneracies \cite{Johannsen:2015hib,Palumbo:2022wnl,EventHorizonTelescope:2020eky,Saurabh:2022oho,Yfantis:2024eqc,Xie:2026ioo,She:2026ProcaHair}.
For theory-driven tests of gravity, implementing this strategy requires an efficient connection between numerical shadow calculations and multidimensional Bayesian inference.
To this end, in Ref. \cite{Liu:2025wwq} we developed a numerically efficient framework that couples ray-traced shadow observables with Markov chain Monte Carlo (MCMC) sampling to jointly infer black-hole and viewing parameters alongside those governing additional gravitational degrees of freedom.
The framework yields marginalized posterior distributions rather than bounds derived from fixed slices of parameter space, thereby providing a practical link between forward shadow modeling and statistically interpretable constraints from horizon-scale observations.

The present problem, however, requires a further step. 
\scopeedit{Even in a multidimensional Bayesian analysis, global marginalization may obscure differences in parameter sensitivity across complex non-Kerr shadow families.}
A shadow boundary may evolve continuously from a nearly elliptical profile to an elongated, compressed, or droplet-like configuration \cite{Liu:2026khl}. 
Along this evolution, however, its parameter sensitivities and associated degeneracy structure may change qualitatively. 
Continuous shadow evolution may therefore mask qualitatively different parameter-response structures and hence distinct inference regimes. 
A global posterior may then combine configurations belonging to different response regimes, obscuring sector-specific degeneracies and complicating the interpretation of marginalized constraints.
In this work, we investigate this problem using an exact rotating non-Kerr spacetime whose nonvanishing radial–angular metric component couples the radial and angular photon motion and obstructs the standard Hamilton–Jacobi separation of Kerr–Newman geodesics \cite{Carter:1968rr}. 
Boundary observables applicable to strongly noncircular contours reveal three morphology-dependent response sectors within the parameter range considered. 
A sector-resolved Bayesian analysis using published EHT constraints for M87* yields distinct parameter degeneracies, motivating separate treatment of the three morphology-response sectors rather than a single globally uniform fit.

\section{Theoretical background}

The inference problem considered here is realized by an exact Kerr-Newman-like spacetime with a nonvanishing radial-angular metric component. 
The solution arises in charged Einstein-bumblebee gravity and provides a controlled setting in which cross-coupled photon dynamics can be connected to shadow morphology and parameter response. 
We first summarize its theoretical origin and then identify the geometric structure relevant to ray tracing and inference.
The underlying field theory contains a vector field $B_\mu$, known as the bumblebee field, whose nonzero vacuum expectation value spontaneously breaks local Lorentz symmetry \cite{Kostelecky2001,Kostelecky:2003fs,Liu:2024wpa,Deng:2025uvp,Lai:2025nyo,Chen:2026zkh,Liu:2026cxs}. 
Because the exact solution considered below is charged, it requires an electromagnetic sector, whose total action is  \cite{Ovcharenko:2026knlv}
\begin{equation}
\begin{aligned}\label{Action}
\!\!\mathcal{S}=&\int d^4x \sqrt{-g}\,
\bigg[
\frac{R}{2\kappa}
+\frac{\varrho}{2\kappa} B^\mu B^\nu R_{\mu\nu}
-\frac{1}{4}B^{\mu\nu}B_{\mu\nu}
-V(X)\\
&\!+\! \bigg(\!-\!\frac14\!+\!\frac{\gamma_1}{2\kappa}B^\lambda B_\lambda\bigg)F_{\mu\nu}F^{\mu\nu}
\!+\!\frac{\gamma_2\!+\!\gamma_3B^\lambda B_\lambda}{2\kappa}
B_\mu F^{\mu\nu}B^\rho F_{\rho\nu}
\bigg],
\end{aligned}
\end{equation}
where $ \kappa$ is the gravitational coupling constant, $A_\mu$ is the electromagnetic four-potential, $F_{\mu\nu}\!=\!\nabla_\mu A_\nu\!-\!\nabla_\nu A_\mu$ is the electromagnetic field strength, and $ \varrho$, $\gamma_1$, $\gamma_2$, and $\gamma_3$ are coupling constants.
The potential $V(X)$, with $X\!=\!B^\mu B_\mu\pm b^2$, induces a nonzero vacuum expectation value $\langle B_\mu\rangle\!\equiv\! b_\mu$ and spontaneous Lorentz-symmetry breaking.
Varying the action with respect to $g_{\mu\nu}$, $B_\mu$, and $A_\mu$ yields the gravitational, bumblebee-field, and modified Maxwell equations; their explicit forms for the charged theory \meq{Action} are given in Ref. \cite{Ovcharenko:2026knlv}.
The role of this field theory in the present analysis is to provide an exact and physically motivated realization of the geometry.

Earlier studies considered exact static solutions \cite{Casana2018,Filho:2022yrk,Chen:2025ypx,Liu:2025oho,Yang:2026jey}, numerical spherical configurations \cite{Xu:2022frb,Ji:2024aeg,Bailey:2025oun,Liu:2025jjf,Luo:2026oxw}, effective photon metrics \cite{Chen2020}, and slowly rotating geometries \cite{Liu:2024axg,Maluf:2022knd,Liu:2022dcn}.
In the static configurations cited above, the Lorentz-violating modification produces no resolved change in the shadow contour relative to the Schwarzschild case, while the effective and slow-rotation treatments do not retain the full higher-order spin-charge structure relevant here.
An exact axisymmetric solution is therefore particularly useful for separating effects associated with rotation, charge, and radial-angular coupling.
Poulis and Soares constructed an exact Kerr-like rotating solution \cite{Poulis:2021nqh}, and Ovcharenko subsequently extended the generating construction to charged electrovacuum spacetimes, yielding Kerr-Newman-like solution extensions \cite{Ovcharenko:2026knlv}, as
\begin{equation}\label{ds2}
\begin{aligned}
ds^2=&-\frac{\Delta}{\Sigma}\left(dt-a\sin^2\theta d\varphi\right)^2
\!+\! \frac{\sin^2\theta}{\Sigma}\left[a dt\!-\!(r^2\!+\!a^2)d\varphi\right]^2\\
&+\frac{\Sigma }{\Delta}dr^2+\Sigma\, d\theta^2
+\frac{\ell}{\Delta} \Big(rdr\pm a\sqrt{\Delta}\cos\theta d\theta \Big)^2.
\end{aligned}
\end{equation}
Here, $\ell\equiv \varrho b^2$ represents the Lorentz violation parameter.
The metric functions $\Delta=r^2+a^2+Q^2-2Mr$ and $\Sigma=r^2+a^2\cos^2\theta$ retain the standard Kerr–Newman form \cite{Newman:1965my}, with $Q$ denoting the electric charge. 
In the charged case, the electromagnetic sector is not described by the minimally coupled Maxwell term alone; instead, the exact generating construction fixes the couplings in Eq. (\ref{Action}) as $\gamma_1=\frac{\kappa\varrho}{2(2+3\ell)}$, $\gamma_2=-\kappa\varrho$, and $\gamma_3=2\varrho\gamma_1$. 
The exact solution considered here provides a charged extension of the Poulis–Soares rotating solution; it reduces to the latter in the limit $Q=0$  \cite{Poulis:2021nqh} and recovers the Kerr–Newman spacetime when $\ell=0$ \cite{Newman:1965my}.
The two signs in \eq{ds2} label two branches of the generated solution. 
They originate from the two possible orientations of the bumblebee one-form in the $(r,\theta)$ sector, or equivalently from the independent choices of the signs of the radial and angular derivatives of the Hamilton-Jacobi generating function. 
After the square in the metric is expanded, the diagonal components are the same for both branches, whereas the off-diagonal component changes sign, $g_{r\theta}^{(\pm)}\!=\!\pm \ell a r\cos\theta \sqrt{1/\Delta} $.
Thus the $(+)$ and $(-)$ branches are not distinguished by their horizons or static limits, but by the orientation of the Lorentz-violating radial-angular coupling.
This distinction becomes relevant only for rotating configurations with $a\ell\neq0$ and controls how the shadows respond to the reversal $a\rightarrow -a$.

The roots of $\Delta$ coincide with those of the Kerr-Newman seed, but this coincidence does not establish the regularity of the corresponding surfaces in the deformed geometry. 
Recent work has identified axial curvature singularities and horizon-regularity obstructions in the related Kerr-disformal family \cite{GuoFan:2026AxialObstructions}, while the charged construction also has limitations concerning the global reality of the bumblebee field \cite{Ovcharenko:2026knlv}. 
We therefore treat the metric as a prescribed exterior geometry and study its optical response subject to an absorbing inner boundary approaching the outer root of $\Delta$. 
The resulting capture contours and parameter constraints are conditional on this boundary prescription and do not establish a globally regular black-hole interpretation.
The same metric also clarifies why previous slow-rotation studies often imposed the additional condition $ \ell^2a^2 \approx 0 $ \cite{Liu:2022dcn,Liu:2024axg}. 
The $d\theta$ component of the bumblebee field generates the cross term $g_{r\theta}$, whose leading coefficient is proportional to $\ell a$.
If this angular component is omitted, the resulting geometry can describe only the first-order slow-rotation sector and misses higher-order rotational corrections induced by Lorentz violation.
The nonzero $g_{r\theta}$ is therefore the key new geometric ingredient of the exact rotating solution.
Our main \scopeedit{focus here} is on the impact on the black hole shadow.

\section{Forward modeling of black hole shadows}

\subsection{ Ray tracing method}

When a black hole is located between the observer and the extended background source, not all photons emitted by the source can reach the observer after being deflected by the black hole gravitational field. 
The apparent shape of the black hole is represented by its shadow contour, which is closely related to the geodesics of photons within the black hole spacetime.
Photon trajectories obey the Hamiltonian constraint, equivalently the Hamilton-Jacobi equation:
\begin{align}\label{EqHJ}
\frac{\pp \mathcal{S}}{\pp \tau}=-\frac{1}{2}g^{ab}\frac{\pp \mathcal{S}}{\pp x^a}\frac{\pp \mathcal{S}}{\pp x^b},
\end{align}
where $ \tau $ is an affine parameter along the null geodesics, and $ \mathcal{S} $ is the Jacobi action. 
We can set up the Killing fields $ \xi^a=(\pp/\pp t)^a $ and $ \psi^a=(\pp/\pp \varphi)^a $, then the corresponding energy $ \mathcal{E} $ and the angular momentum in the direction of the axis of symmetry $ L $ can be defined as:
\begin{align}\label{gabE}
\mathcal{E}=-g_{ab}\xi^a\dot{x}^b,&&
L=g_{ab}\psi^a\dot{x}^b.
\end{align}
Using these two conserved quantities, the null geodesics for the metric \meq{ds2} can be expressed as follows:
\begin{equation}
\begin{aligned}
&\begin{aligned}\label{tauvarphi}
\Sigma\dot{t}=&{\big(r^2+a^2\big)^2\mathcal{E}/\Delta
-a\big(2Mr-Q^2\big)L/\Delta}
-a^2\mathcal{E}\sin^2\theta, 
\end{aligned}\\
&\begin{aligned}
\Sigma\dot{\varphi}=&{a\big(2Mr-Q^2\big)\mathcal{E}/\Delta-a^2L/\Delta}
+L\csc^2\theta,
\end{aligned}\\
&\begin{aligned}
\Sigma\big(g_{rr}\dot{r}^2+g_{\theta\theta}\dot{\theta}^2
+2 g_{r\theta}^{(\pm)}\dot{r}\dot{\theta}\big)=&
\big[(r^2+a^2)\mathcal{E}-aL\big]^2/\Delta\\
&-\left(a\,\mathcal{E}\sin\theta-L\csc\theta\right)^2.
\end{aligned}
\end{aligned}
\end{equation}
With these equations, we can study the propagation of light in the Kerr-Newman-like Lorentz-violating black hole spacetime and probe the properties of the black hole shadow cast by the light.
It is evident that the bumblebee field introduces cross terms, $ \dot{r}\dot{\theta} $, i.e., the $r$-component and $ \theta $-component do not decouple from each other in rotating spacetime. 
This obstructs the usual Kerr-Newman-type separation of the radial and angular equations by a Carter constant.
For this reason, we solve the null geodesic equations numerically and determine the component values of the 4-momentum directly.
For this, we can adopt the ``backward ray-tracing" method \cite{Cunha:2015yba,Cunha:2019dwb,Wang:2017qhh,Hu:2020usx,Zhong:2021mty,Garnier:2023lph,Huang:2024wpj,Zhang:2025xnl} to obtain the shadow cast by the Kerr-Newman-like Lorentz-violating black hole.
In this computational construction, light rays are launched from the observer into the scene, with each initial direction mapped to a specific pixel on the observer's screen.

For the black hole spacetime $ g_{\mu\nu} $ and the locally observed spacetime $ \eta_{ab} $, the following transformation relationship holds, $ g^{\mu\nu}=\eta^{ab} e_a^{~\mu}e_b^{~\nu} $, where $ e_a^{~\mu} $ is the transform matrix.
The observer is at $(r_0,\theta_0)$, where $r_0$ is the radial coordinate and $\theta_0$ is the inclination relative to the rotation axis.
By solving this relationship, we can obtain the transform matrix $ e_{a}^{~\mu} $, and the orthogonal tetrad form $ e_{(a)}=e_{a}^{~\mu}\pp_\mu $, as 
\begin{equation}\label{zjbj}
\begin{aligned}
e_{(t)}=&\left.\sqrt{\frac{g_{\varphi\varphi}}{g_{t\varphi}^2-g_{tt}g_{\varphi\varphi}}}\left(\pp_t-\frac{g_{t\varphi}}{g_{\varphi\varphi}}\pp_\varphi\right)\right|_{(r_0,\theta_0)},\\
e_{(r)}=&\left.      \sqrt{\frac{g_{\theta\theta}}{g_{rr}g_{\theta\theta}-g_{r\theta}^2}} 
\left( -\pp_r+\frac{g_{r\theta}}{g_{\theta\theta}}\pp_\theta \right)    \right|_{(r_0,\theta_0)},\\
e_{(\theta)}=&\left.\frac{1}{\sqrt{g_{\theta\theta}}}\pp_\theta\right|_{(r_0,\theta_0)},\\
e_{(\varphi)}=&\left.\frac{1}{\sqrt{g_{\varphi\varphi}}}\pp_\varphi\right|_{(r_0,\theta_0)}.
\end{aligned}
\end{equation}
where $ g_{\mu\nu} $ is the black hole metric component of the background. 
Here we want to stress that this tetrad is not the only choice: one can select an appropriate tetrad based on actual requirements.
Using the orthogonal tetrad, the four-momentum can be expressed as $ p^{(t)}=-p_\mu e^{\mu}_{(t)} $ and $ p^{(i)}=p_\mu e^{\mu}_{(i)} $, where $ i=r,\theta,\varphi $.
This represents the four-momentum as measured by a \scopeedit{locally nonrotating} observer.
Since the photon is massless, its three-vector linear momentum $\vec{P}$ is related to $p^{(i)}$ and satisfies $|\vec{P}| = p^{(t)}$ in the observer's frame.
The observation angles $ (\alpha, \beta) $ can be defined in the following:
\begin{equation}\label{obsangles}
\begin{aligned}
p^{(r)}&=|\vec{P}|\cos\alpha \cos\beta,\\
p^{(\theta)}&=|\vec{P}|\sin\alpha,\\
p^{(\varphi)}&=|\vec{P}|\cos\alpha\sin\beta.
\end{aligned}
\end{equation}
Actually, the angular coordinates $ (\alpha, \beta) $ of a point in the observer's local sky determine the direction of the corresponding light ray and set its initial conditions.
The coordinates $(X, Y)$ of a point in the observer's local sky are related to its angular coordinates $(\alpha, \beta)$ as follows:
\begin{equation}\label{screenxy}
\begin{aligned}
X=-r_0\tan\beta=-r_0\frac{p^{(\varphi)}}{p^{(r)}},&&
Y=r_0\frac{\tan\alpha}{\cos\beta}=r_0\frac{p^{(\theta)}}{p^{(r)}}.
\end{aligned}
\end{equation}
We define the shadow operationally as the set of screen directions whose backward-traced rays reach the prescribed absorbing inner boundary before escaping to the celestial sphere. 
This definition characterizes the capture image of the exterior model and does not assume a regular extension across the inner boundary.

\begin{figure}[h]
\centering
\includegraphics[width=0.85\linewidth]{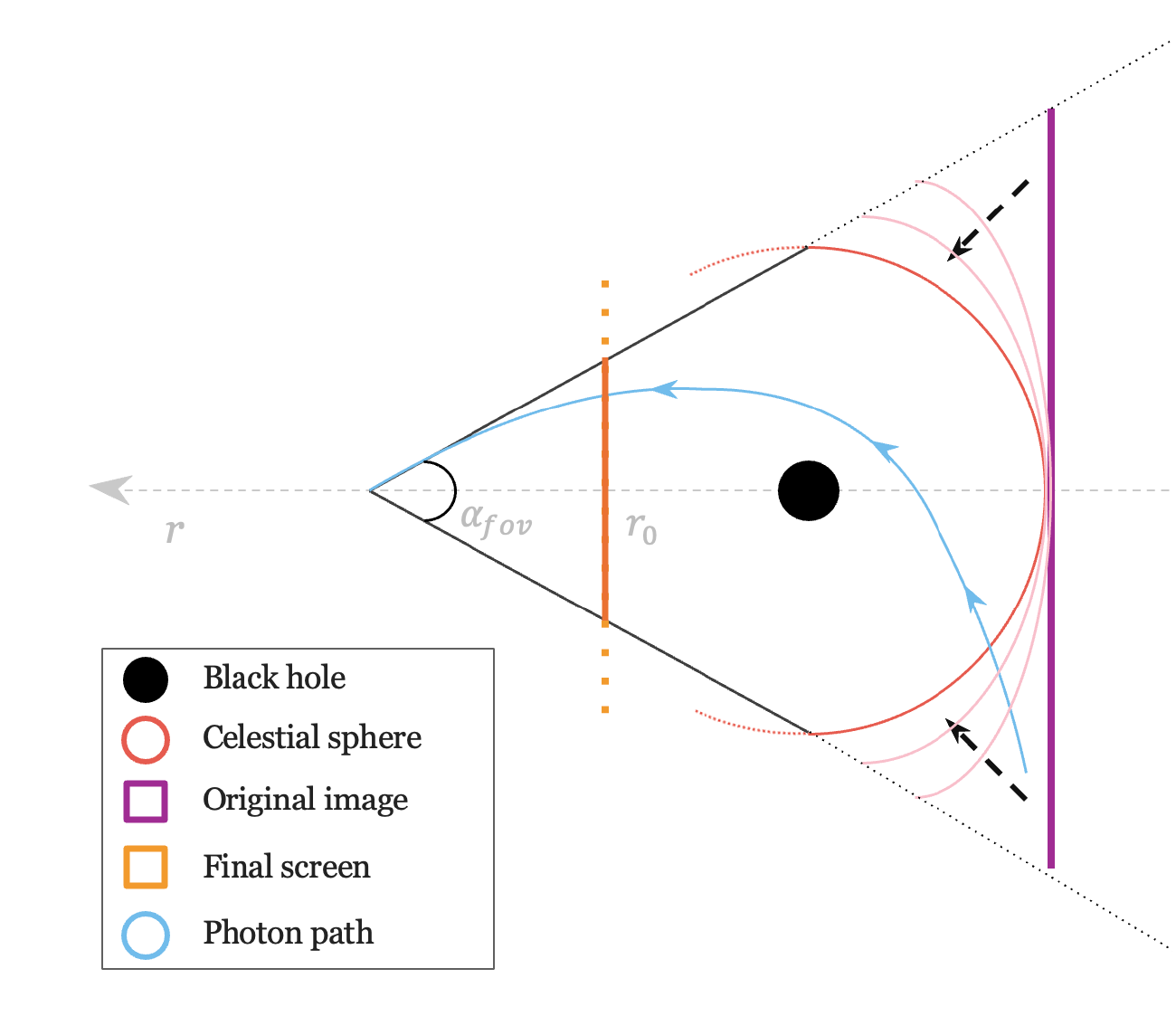}
\caption{Schematic relation between the original background image and the final image on the observer's screen.
The exterior geometry lenses the background source onto the screen; directions captured under the absorbing inner-boundary prescription define the dark region.}
\label{fig1}
\end{figure}

\begin{figure}[h]
\centering
\includegraphics[width=0.68\linewidth]{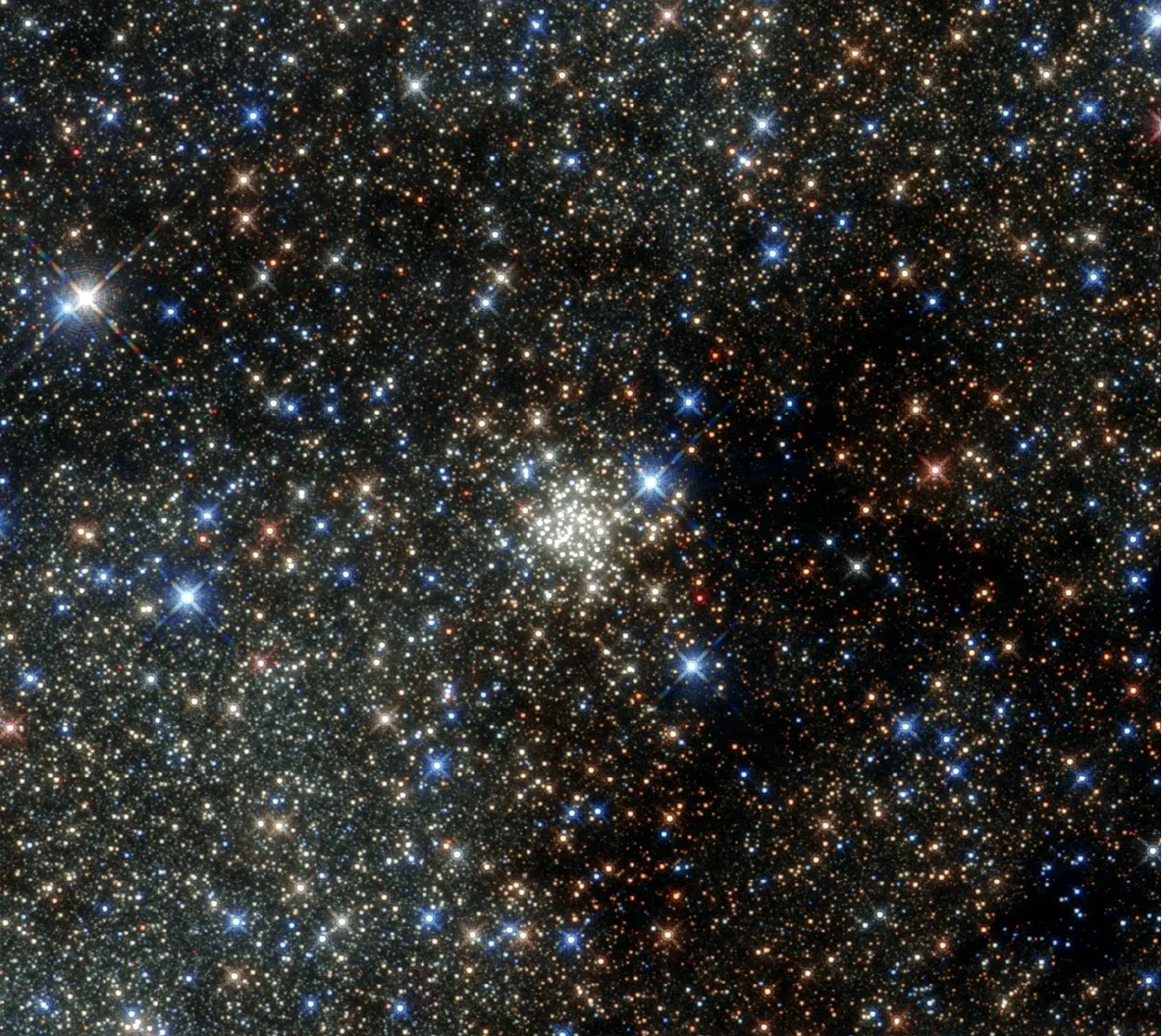}
\caption{Background image used for the ray-tracing visualization.
Image credit: NASA/ESA; source: NASA Science~\cite{NASAArchesCluster}.}
\label{fig2}
\end{figure}

We next specify the observational configuration used to produce the simulated images.
As illustrated in Fig. \ref{fig1}, an original background image is projected onto a distant celestial sphere, and the observed image is recorded on the local screen of an observer at $(r_0,\theta_0)$ with angular field of view $\alpha_{\rm fov}$.
The black hole lies between these two image planes and maps the background source onto the observer's screen through gravitational lensing.
In all image comparisons below, the observer position, field of view, screen resolution, and outer celestial sphere are kept fixed, so that changes in the final image arise only from the black hole parameters.
The inclination $\theta_0$ is not an intrinsic black hole parameter such as $M$, $a$, $Q$, or $\ell$, but it becomes an important observational parameter for rotating black holes because the loss of spherical symmetry makes the projected shadow and lensing pattern depend on the viewing inclination.
In the Bayesian inference analysis of rotating black holes in Sec. \ref{sec3}, $\theta_0$ should therefore be inferred together with the spacetime and Lorentz-violation parameters.
Unless otherwise stated, the observer is placed in the equatorial plane, $\theta_0=\pi/2$, in the examples below.

For each pixel $(X,Y)$ on the screen, the local angles $(\alpha,\beta)$ are obtained from \eq{screenxy}, and the corresponding photon momentum in the observer's tetrad is fixed by \eq{obsangles}.
The tetrad \eq{zjbj} then converts this local momentum into coordinate components, which are used as initial data for the null geodesic equations.
The capture-versus-escape construction is interpreted here as ray tracing through the prescribed exterior geometry with an absorbing inner boundary near $r_+$. Rays assigned to this inner boundary are represented by black pixels in the capture image.
If the ray reaches the celestial sphere, the pixel color is assigned from the intersection point on the background image.
Thus the shadow contour is determined by capture versus escape, while the surrounding image records the gravitationally lensed distortion of the extended background source.
For the lensing visualization, the astronomical background shown in Fig. \ref{fig2} is taken from the NASA/ESA Hubble Space Telescope image of the Arches Cluster \cite{NASAArchesCluster}.
This dense stellar field provides an extended source for displaying the detailed lensing morphology, whereas the schematic setting in Fig. \ref{fig1} clarifies how the original image is mapped into the final screen image.

\subsection{Black hole shadow and gravitational lensing}

\begin{figure}[th]
\centering
\includegraphics[width=1\linewidth]{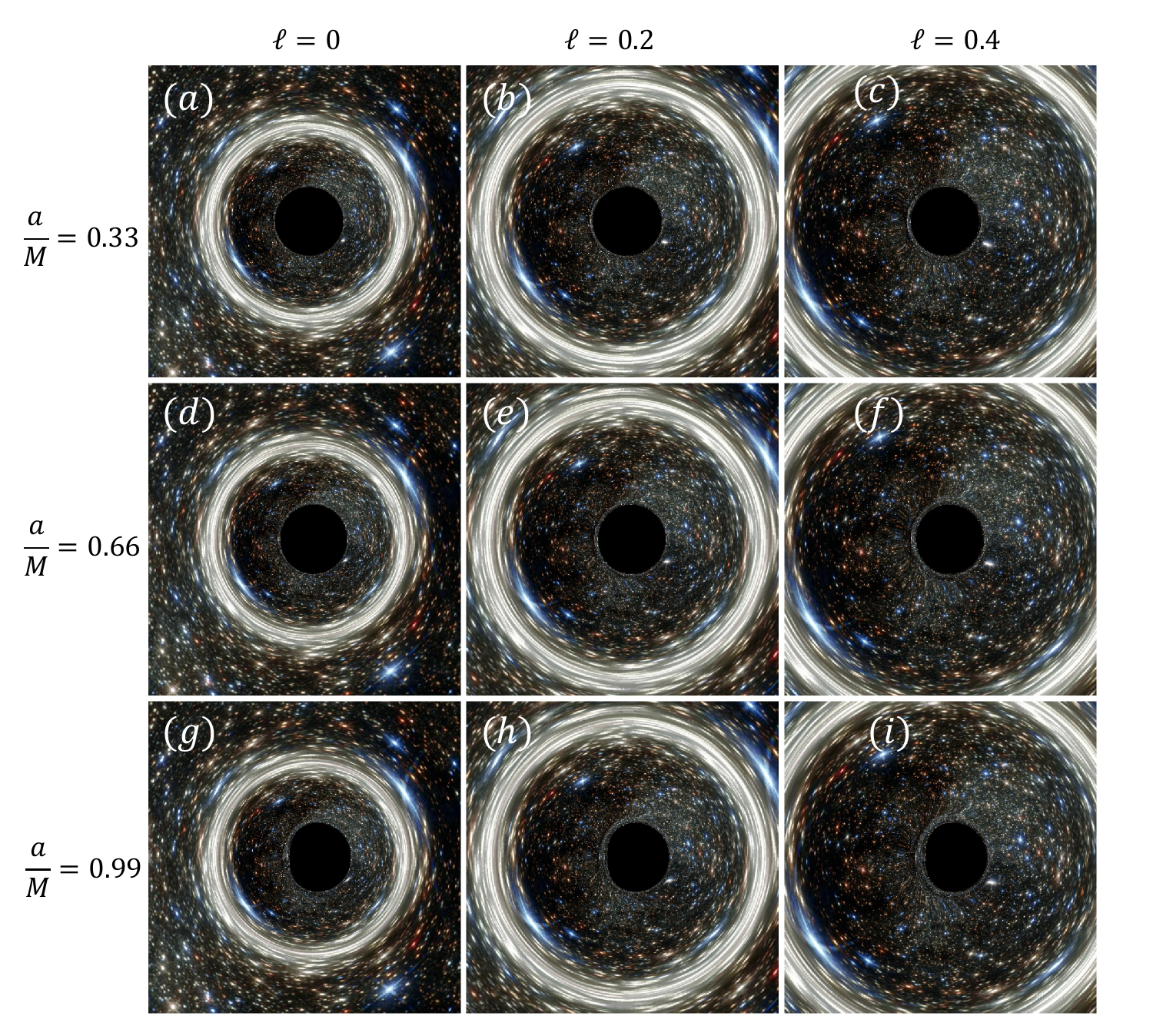}
\caption{Kerr-like black hole shadows in the uncharged case $Q=0$ with a positive Lorentz violation parameter $\ell$, as seen by an observer at $\theta_0=\pi/2$. The Arches Cluster image in Fig.~\ref{fig2} is used as the background for ray tracing.}
\label{fig3}
\end{figure}

In this subsection, we show the black hole shadow and gravitational lensing under the observational configuration $r_0 = 40M, \alpha_{\rm fov}=\pi/3$. 
We first consider the uncharged case $Q=0$, which provides the Kerr-like limit of the exact Kerr-Newman-like Lorentz-violating solution.
For simplicity, we focus on the branch where $ (\pm) $ is $ (+) $ in metric \meq{ds2}.
Figs. \ref{fig3}(a)-\ref{fig3}(i) describe the cases with a positive Lorentz violation parameter.
When the parameter $a/M=0$, the geometry reduces to the Schwarzschild-like bumblebee branch.
In this static uncharged limit, the region affected by gravitational lensing changes with the Lorentz violation parameter, but the shadow contour itself remains the same as that of the Schwarzschild spacetime.
This behavior is expected from the structure of the static limit. 
For the static Lorentz-violating black hole, the radial part of the photon trajectory can be obtained by defining the Carter constant. 
Its effective potential is independent of the metric function $g_{rr}$, while the Lorentz violation correction affects precisely $g_{rr}$.
When $a/M \neq 0$, positive Lorentz violation weakens the \scopeedit{rotational distortion of the lensing pattern}. 
This can be seen directly from identifiable structures in the background image. 
For example, the localized bright feature near the lower-right side of the shadow and the surrounding stellar texture are increasingly sheared as the spin grows, whereas this rotational distortion becomes less pronounced when positive $\ell$ is increased. 
When $\ell$ is large, the gravitational lensing edge region of a rapidly rotating black hole therefore approaches the static pattern. 
Regarding the shadow itself, positive Lorentz violation does not produce a significant increase or decrease in the contour size within the parameter range shown here. 
\begin{figure}[th]
\centering
\includegraphics[width=1\linewidth]{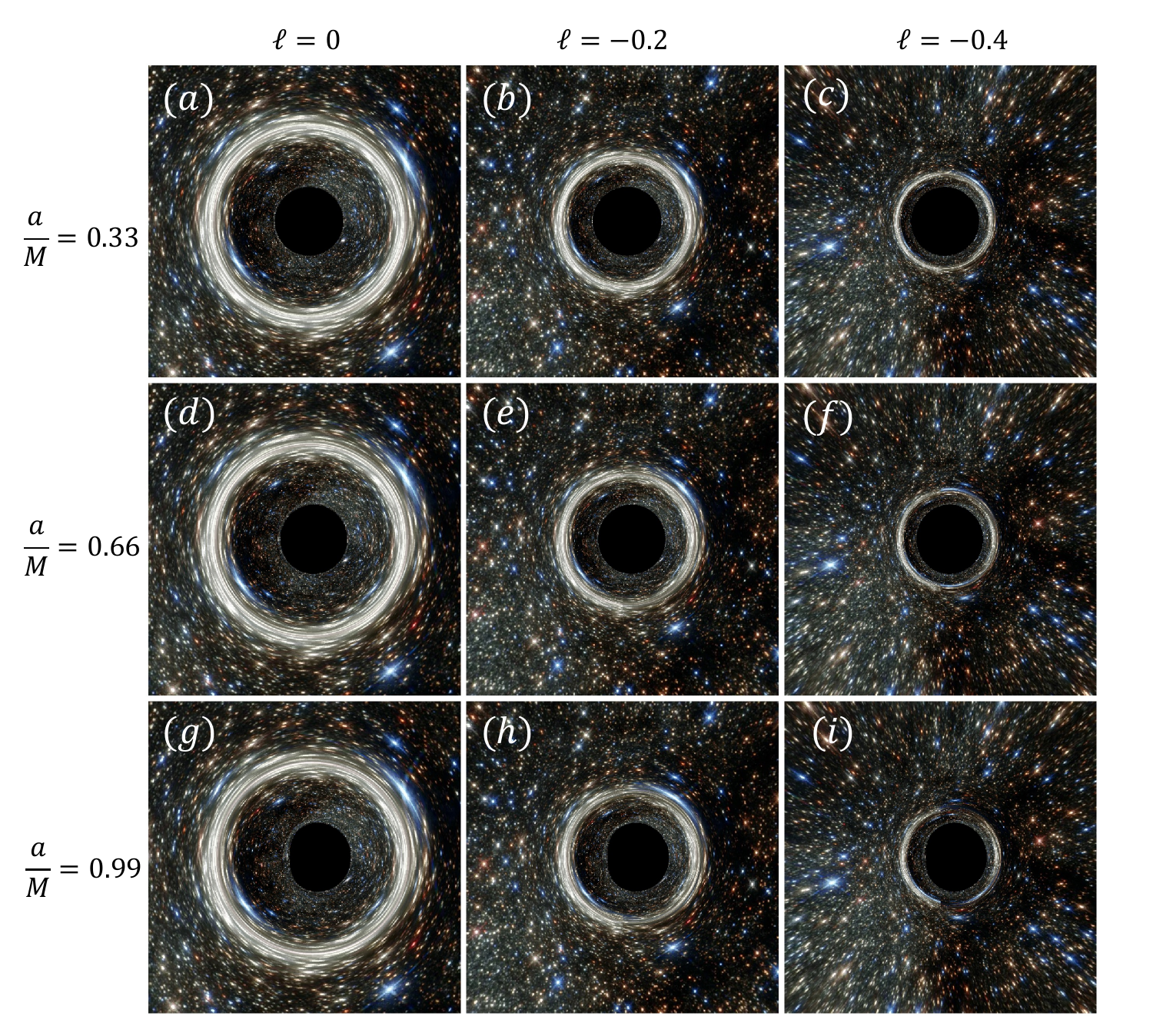}
\caption{Kerr-like black hole shadows in the uncharged case $Q=0$ with a negative Lorentz violation parameter $\ell$, as seen by an observer at $\theta_0=\pi/2$. The Arches Cluster image in Fig.~\ref{fig2} is used as the background for ray tracing.}
\label{fig4}
\end{figure}
For small spin parameters, the shadow contours for different values of $\ell$ are almost indistinguishable. 
This behavior is qualitatively consistent with the result of Liu et al. for slowly rotating Kalb-Ramond black holes \cite{Liu:2024lve}, where it was shown analytically that the Lorentz-violating parameter does not modify the shadow contour at first order in the rotation parameter. 
In the rapidly rotating case considered here, the images indicate that positive $\ell$ mainly changes the lensing pattern rather than producing a clearly resolvable deformation of the shadow contour.



\begin{figure}[th]
\centering
\includegraphics[width=1\linewidth]{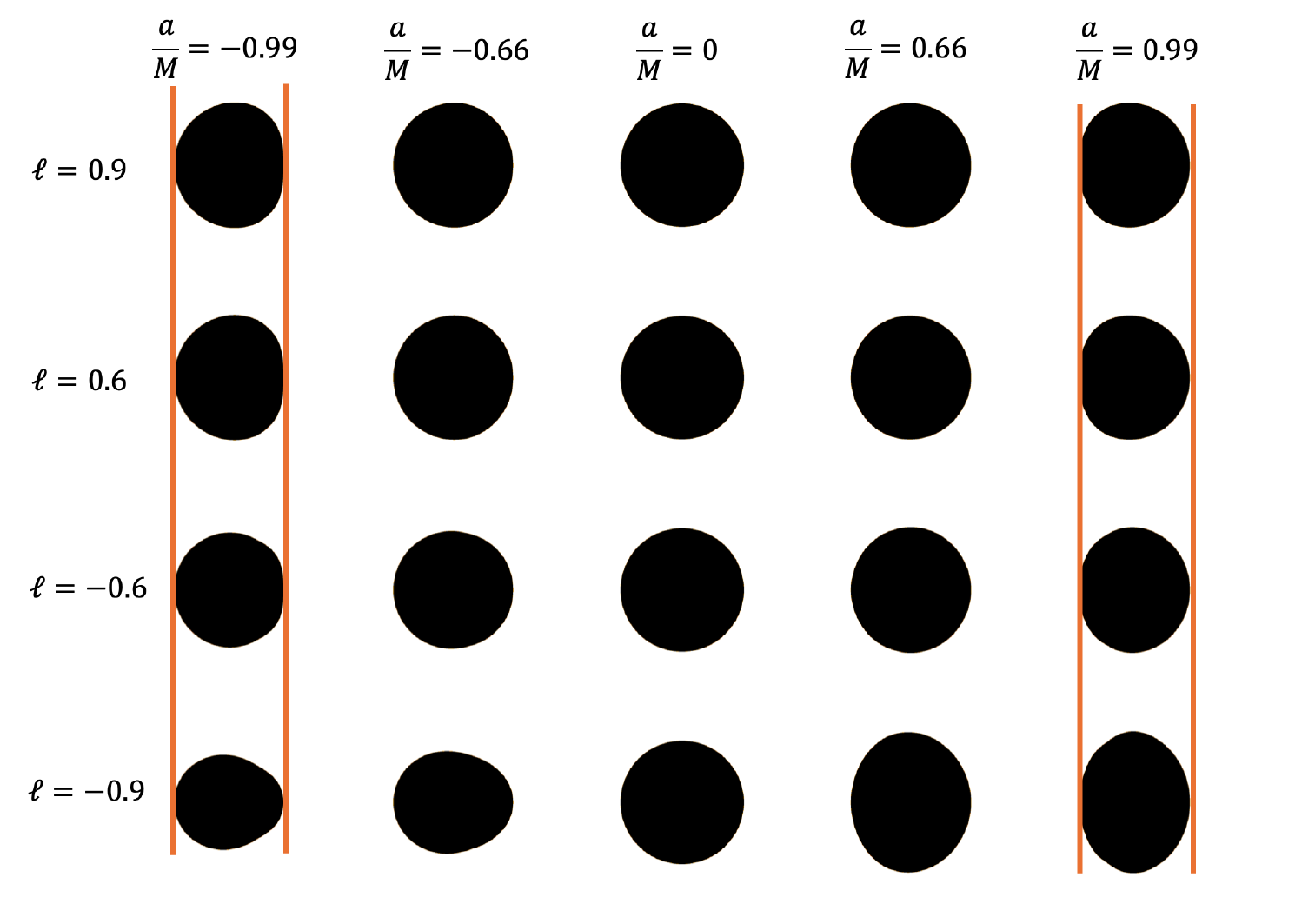}
\caption{Shadow contours cast by Kerr-like black holes in the uncharged case $Q=0$ for different spin parameters and Lorentz violation parameters, as seen by an observer at $\theta_0=\pi/2$. Red vertical lines in the outermost columns serve as guides for comparing the horizontal shadow width across rows.}
\label{fig5}
\end{figure}

Therefore, we will next discuss the impact of the negative Lorentz violation parameter on the shadow contours. 
Similarly, in the static uncharged case, Lorentz violation only affects gravitational lensing.
Figs. \ref{fig4}(a)-\ref{fig4}(c) can be seen as the inverse continuation of Figs. \ref{fig3}(a)-\ref{fig3}(c), where the lensing pattern changes while the shadow boundary is still fixed by the Schwarzschild geometry.
For rotating black holes shown in Figs. \ref{fig4}(d)-\ref{fig4}(i), negative Lorentz violation \scopeedit{enhances the rotational distortion of the lensing pattern}. 
Interestingly, for near-extreme black holes, the shadow contour changes \scopeedit{as $\ell$ becomes more negative}. 
The D-shaped structure of the near-extreme Kerr shadow is gradually weakened and tends toward a more rounded contour, as shown in Fig. \ref{fig4}(i).
This also implies that for rotating black holes, the non-minimally coupled bumblebee field is not completely decoupled from the shadow contour, at least in the higher-order rotational terms, where modifications will arise.

\begin{figure*}[th]
\centering
\includegraphics[width=0.8\linewidth]{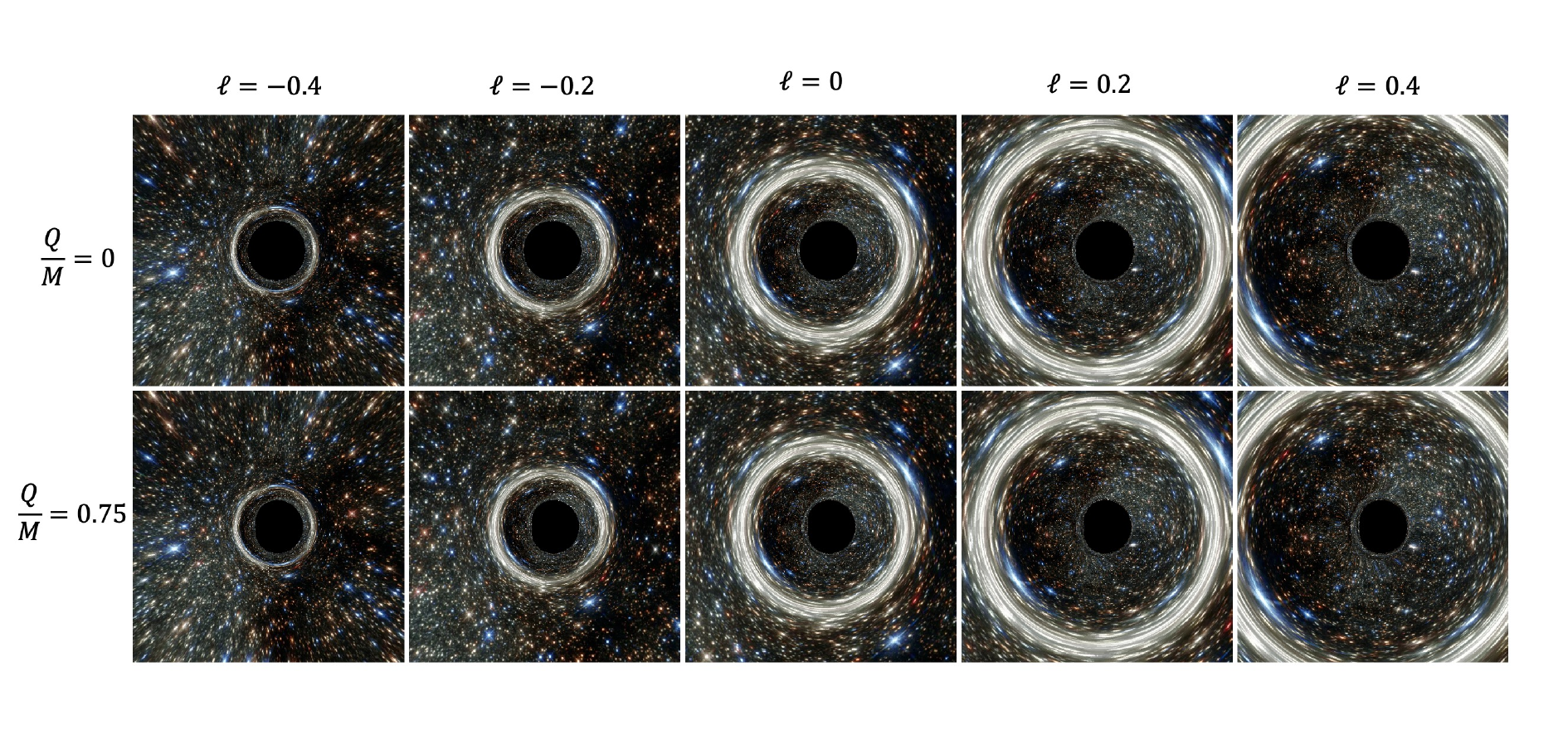}
\caption{Black hole shadows and gravitational lensing images for the Kerr-Newman-like Lorentz-violating black hole at fixed spin $a/M=0.66$, as seen by an observer at $\theta_0=\pi/2$. 
The upper row shows the uncharged case $Q/M=0$, while the lower row shows the charged case $Q/M=0.75$. 
From left to right, the Lorentz violation parameter is $\ell=-0.4,-0.2,0,0.2,0.4$.}
\label{fig6}
\end{figure*}
The full images in Figs. \ref{fig3} and \ref{fig4} show that Lorentz violation can strongly modify the lensing pattern, while its influence on the shadow boundary is more subtle. 
To isolate this weaker effect from the background distortion, we now remove the celestial-sphere image and compare only the shadow contours. 
This contour-only comparison is also useful for highlighting the effect of opposite spin directions, where the branch-dependent radial-angular coupling can make the deformation of the shadow shape more visible.
Fig. \ref{fig5} shows a collection of uncharged Lorentz-violating black hole shadow contours formed by different spin parameters and Lorentz violation parameters. 
It demonstrates the potential impact of the spontaneous Lorentz-violating field coupled to gravity. 
A notable feature is that the shadow contours of counter-rotating Lorentz-violating black holes are not simply mirror images of those of co-rotating ones, unlike the corresponding Kerr case in general relativity. 
This asymmetry is naturally associated with the branch-dependent radial-angular coupling $g_{r\theta}$ and with the anisotropic effective source $T^{B}_{\mu\nu}$ \cite{Ovcharenko:2026knlv}.
We focus on the shadow images in the leftmost and rightmost columns of Fig. \ref{fig5}. 
With the help of auxiliary lines, we can determine that the Lorentz violation parameter has little effect on the horizontal width of the shadow, i.e., the position information on the X-axis in the Cartesian coordinate system. 
\scopeedit{As $\ell$ becomes more negative}, the height of the counter-rotating black hole shadow is compressed, gradually transforming from a reverse D-shape to a droplet shape. 
In contrast, the height of the co-rotating black hole shadow is stretched, and the D-shaped structure gradually transforms into an elliptical shape.
Now we have a clearer understanding of the impact of Lorentz violation: it has little effect on the horizontal width in the parameter ranges shown here, but it stretches or compresses the vertical height and thereby changes the shadow shape.

This raises a natural question: why does this asymmetry occur? 
The key point is the relation between the branch sign and the spin sign in the cross term of metric \meq{ds2}. 
For the two branches, one has $g_{r\theta}^{(\pm)}\!=\!\pm {\ell a r\cos\theta}/{\sqrt{\Delta}}$.
Thus the sign of the radial-angular coupling is controlled by the product of the branch sign and the spin parameter $a$. 
Changing only $a\rightarrow -a$ within a fixed branch reverses the Lorentz-violating radial-angular coupling, so the shadows for $a$ and $-a$ in the same branch need not be mirror images. 
By contrast, the simultaneous transformation $(s,a,\varphi)\rightarrow(-s,-a,-\varphi)$, where $s=\pm1$ denotes the branch sign, leaves the line element invariant: the product $sa$ in $g_{r\theta}^{(s)}$ is unchanged, while the sign flip of the Kerr-Newman term $g_{t\varphi}$ is compensated by $\varphi\rightarrow-\varphi$. 
Therefore, the $(+)$ branch with spin $a$ and the $(-)$ branch with spin $-a$ are analytically equivalent up to this azimuthal reflection. 
On the observer's screen, this corresponds to a reflected image rather than to a point-by-point identical image in the same screen coordinates. 
With this qualification, the $(+)$ branch can be used as the representative branch in the image survey above.

The uncharged results above establish the baseline role of the Lorentz violation parameter: in the absence of electric charge, $\ell$ first modifies the lensing pattern and, for rapidly rotating black holes with sufficiently negative values, can also deform the shadow contour. 
We will next turn to the genuinely charged case $Q\neq0$.
This extension is important because the electric charge changes the Kerr-Newman part of the geometry through $\Delta$, while the Lorentz-violating term introduces an additional radial-angular coupling.
Comparing the charged and uncharged images will allow us to separate the effects of charge from those produced by Lorentz violation.

Figure \ref{fig6} gives a direct comparison between the uncharged and charged cases at the same spin and observational configuration. 
The charged row corresponds to $Q/M\!=\!0.75$ and $a/M\!=\!0.66$, which is close to the Kerr-Newman extremal bound $a^2\!+\!Q^2\!=\!M^2$. 
Therefore, the lower panels probe a genuinely strong-field charged configuration rather than a small perturbation of the Kerr-like case. 
Comparing the two rows column by column, the most direct effect of the electric charge is the reduction of the apparent shadow size, as expected from the Kerr-Newman part of the geometry through the modified radial function $\Delta$. 
By contrast, the change of the lensing ring or high-magnification background structures caused by $Q$ is not visually isolated in this set of images. 
The qualitative dependence on $\ell$ remains nearly parallel in the two rows: negative $\ell$ enhances the rotational distortion of the background image, while positive $\ell$ weakens this distortion and makes the outer lensing pattern closer to the static-like case. 
Thus, at the level of the simulated images in Fig. \ref{fig6}, the charge effect and the Lorentz-violating effect appear to be approximately superposed: $Q$ mainly changes the Kerr-Newman shadow scale, whereas $\ell$ continues to control the rotational deformation of the lensing pattern.
Whether a genuine coupling between $Q$ and $\ell$ exists, however, cannot be determined from visual inspection alone and should be tested through quantitative shadow observables extracted from the simulated ray-tracing data.
In the next subsection, we will discuss the observables used to characterize the non-circular shadow boundary.

\subsection{Observables}\label{subsec:observables}
\begin{figure}[h]
\centering
\includegraphics[width=1\linewidth]{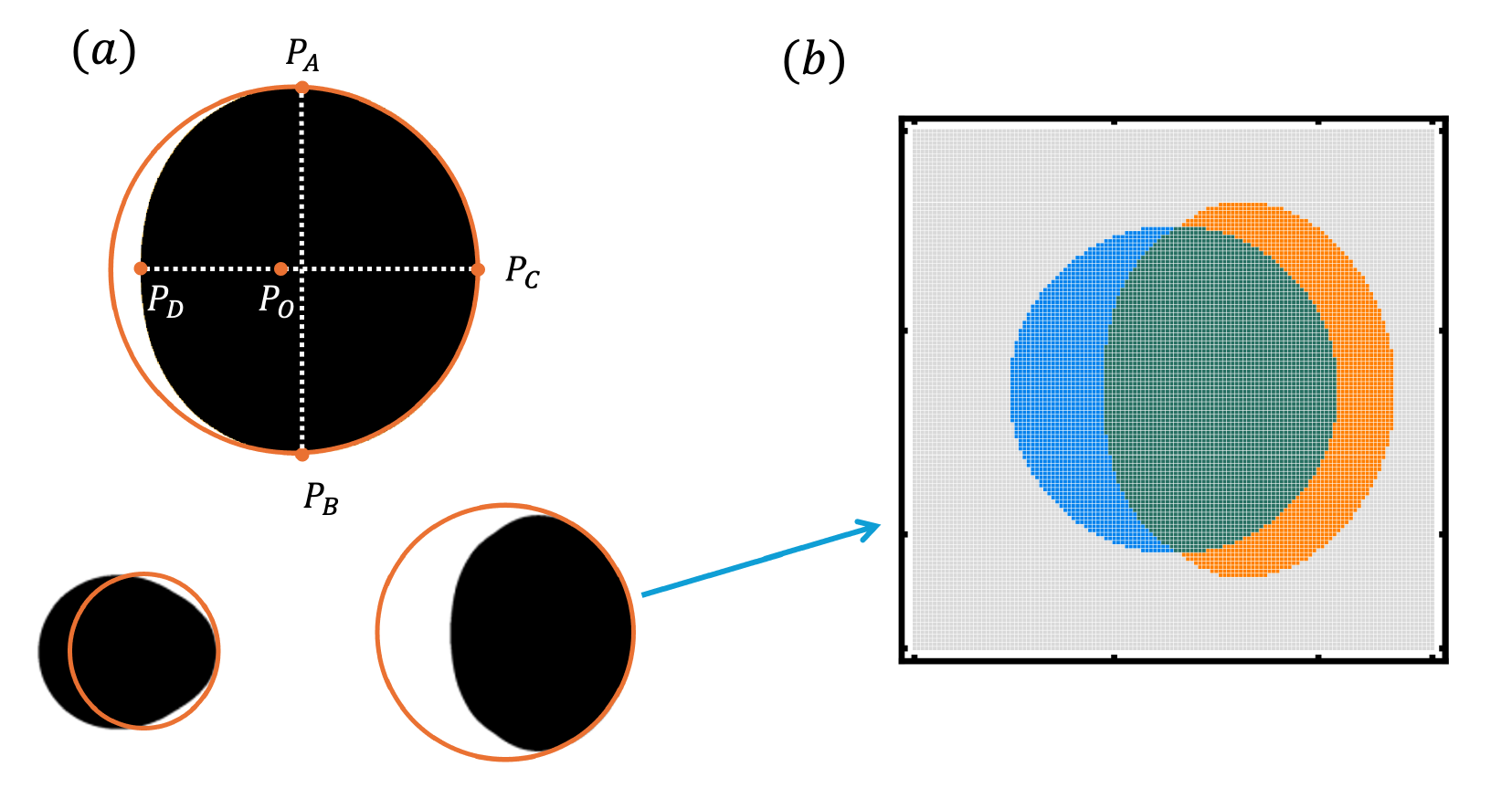}
\caption{(a) Reference-circle construction and representative non-circular shadow examples. 
(b) Pixelized screen-plane comparison between a Schwarzschild shadow and the target black hole shadow under the same observational configuration, used to calibrate the shadow area. Green pixels belong to both shadows, yellow pixels only to the target shadow, and blue pixels only to the Schwarzschild shadow.}
\label{fig7}
\end{figure}
The image comparisons above show that Lorentz-violating rotating black holes can produce highly non-circular shadow boundaries, including droplet-like and elliptical profiles. 
To quantify a shadow image, one usually begins by defining a reference circle \cite{Hioki:2009na}. 
As shown in the upper part of Fig. \ref{fig7}(a), this circle is determined from characteristic boundary points: the top, bottom, and rightmost points $P_A$, $P_B$, and $P_C$ fix the circle, while $P_D$ denotes the leftmost point of the shadow and $P_O$ is the circle center. 
For exotic shadows \cite{Cunha:2016bjh,Cunha:2017eoe,Kuang:2024ugn}, however, a characterization based solely on this reference circle is no longer sufficient, because it can obscure the stretching, compression, and spin-dependent deformation generated in the simulated images. 
The same prescription may also either miss part of the non-circular shadow boundary or enclose a large empty region, as illustrated by the lower examples in Fig. \ref{fig7}(a). 
We therefore retain the reference-circle size as a baseline measure of the shadow scale, but supplement it with additional observables constructed from the shadow boundary, including the longest and shortest axes, the shadow area, and the corresponding axis ratio.

\begin{figure*}[th]
\centering
\includegraphics[width=0.65\linewidth]{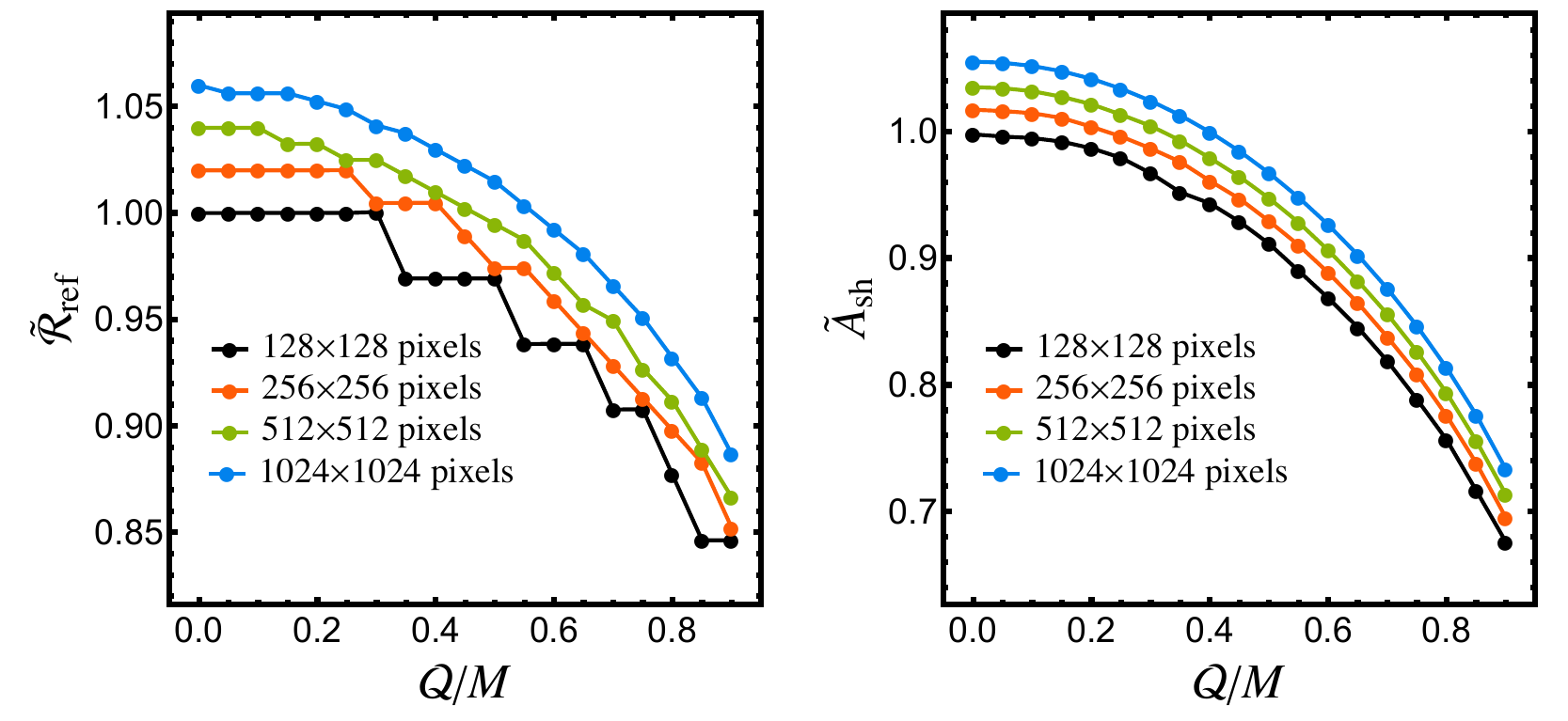}
\caption{Resolution dependence of the normalized reference-circle radius $\tilde{\mathcal{R}}_{\rm ref}$ and the area observable $ \tilde{A}_{\rm sh}$ in the Kerr-Newman limit.
The spin is fixed at $a/M=0.3$, while the charge varies from $Q/M=0$ to $Q/M=0.9 $ with step size 0.05. 
For visual clarity, the results at resolutions (256), (512), and (1024) are shifted upward by (0.02), (0.04), and (0.06), respectively.}
\label{fig8}
\end{figure*}
In the following analysis, we use the reference-circle radius $\mathcal{R}_{\rm ref}$, the shadow area $A_{\rm sh}$, the longest and shortest axes $d_{\max}$ and $d_{\min}$, and the corresponding axis ratio $\mathcal{D}_{\rm ax}$ as shadow observables. 
Let $d_{(i,j)}$ denote the Euclidean distance between two points $P_i$ and $P_j$ on the observer's screen.
The reference-circle radius and the axis observables are then defined as
\begin{equation}
\begin{aligned}
&\, \mathcal{R}_{\rm ref} = d_{(A,O)},\quad\quad\quad\quad\quad\quad
  \mathcal{D}_{\rm ax}={d_{\max}}/{d_{\min}}, \\
&d_{\max} = \max\{d_{(A,B)},d_{(C,D)}\}, \quad
d_{\min} = \min\{d_{(A,B)},d_{(C,D)}\}.
\end{aligned}
\end{equation}
For the area observable, $A_{\rm sh}$ is evaluated directly from the screen image. 
The procedure follows the Schwarzschild calibration idea used in the Supplemental Material of Ref. \cite{Liu:2025wwq}: a Schwarzschild shadow is computed under the same observational configuration and used as a reference standard for the pixel scale. 
The difference is that Ref. \cite{Liu:2025wwq} uses the Schwarzschild shadow-area fraction as a calibration quantity for the image scale, whereas here we use the same pixel-ratio idea to define the physical shadow area itself. 
Let $N_{\rm tar}$ and $N_{\rm Sch}$ be the numbers of shadow pixels for the target black hole and for the Schwarzschild black hole, respectively, evaluated with the same screen resolution, field of view, and observer position. 
In Fig. \ref{fig7}(b), $N_{\rm tar}$ corresponds to the yellow plus green pixels, while $N_{\rm Sch}$ corresponds to the blue plus green pixels. 
Then
\begin{equation}
A_{\rm sh}=\frac{N_{\rm tar}}{N_{\rm Sch}}A_{\rm Sch},
\end{equation}
where $A_{\rm Sch}\!=\!27\pi M^2$ is the known Schwarzschild shadow area. 
This area-evaluation method does not rely on an analytic boundary integral and is applicable to arbitrary stationary black-hole spacetimes. 
Since $A_{\rm sh}$ is obtained from the pixel occupation of the whole shadow region, it is a global observable and is less sensitive to local boundary-point errors. 
In contrast, the reference-circle radius $\mathcal{R}_{\rm ref}$ depends only on a few characteristic points $P_A$, $P_B$, and $P_C$, and is therefore more sensitive to pixel-location errors when the image resolution is low or the shadow boundary is irregular. 
As will be shown in Fig. \ref{fig8}, the area observable converges faster than the reference-circle radius and can already achieve high accuracy at relatively low image resolution.
These quantities provide a quantitative description of the shadow morphology and allow us to track how the forward-modeled image changes with $a$, $Q$, $\ell$, and the viewing angle $\theta_0$. 
In particular, the possible interplay between $Q$ and $\ell$ should be examined through the parameter dependence of the observables extracted from the simulated shadow boundaries, rather than inferred from the visual appearance of the images alone. 
In this sense, the observables are the quantitative output of the forward modeling and provide the data-level basis for the Bayesian analysis in Sec. \ref{sec3}, where the spacetime parameters, Lorentz-violation parameter, and observational parameters will be inferred \scopeedit{by comparing these predictions with EHT constraints}.

\begin{figure*}[th]
\centering
\includegraphics[width=0.9\linewidth]{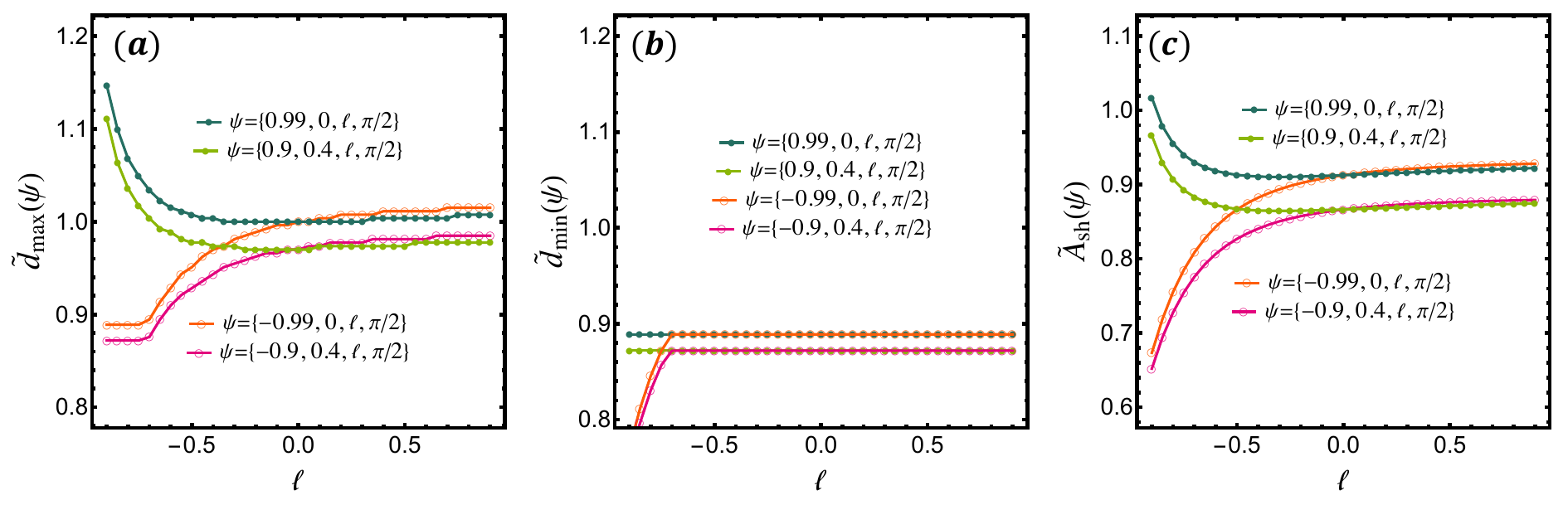}
\caption{Dependence of the normalized shadow observables on the Lorentz-violating parameter $\ell$ for different spin and charge configurations. 
Panels (a), (b), and (c) correspond to the normalized longest axis $\tilde d_{\rm max}$, shortest axis $\tilde d_{\rm min}$, and shadow area $\tilde A_{\rm sh}$, respectively.}
\label{fig9}
\end{figure*}

To quantitatively isolate the structural deformations discussed above, we remove the overall mass scale by normalizing the relevant observables with respect to the Schwarzschild case under the same observational configuration.
We therefore introduce the dimensionless observables
\begin{equation}\label{eq:normalized_observables}
 \tilde{\mathcal{R}}_{\rm ref}\!=\! \frac{\mathcal{R}_{\rm ref}(\psi)}{\mathcal{R}_{\rm ref}(\psi_{\rm Sch})},
\quad
\tilde{{A}}_{\rm sh}\!=\! \frac{{A}_{\rm sh}(\psi)}{{A}_{\rm sh}(\psi_{\rm Sch})},
\quad
\tilde{d} \!=\! \frac{{d}(\psi)}{{d}(\psi_{\rm Sch})}.
\end{equation}
Here $\psi \!=\! \{ \tilde{a},\tilde{Q},\ell,\theta_0 \} $, where $\tilde{a} \!=\! \frac{a}{M} $ and $ \tilde{Q} \!=\! \frac{Q}{M} $ denote the dimensionless spin and charge, respectively.
The symbol $\psi_{\rm Sch} \!=\! \{ 0,0,0,\theta_0 \} $ denotes the Schwarzschild reference configuration evaluated with the same observer position, field of view, and image resolution.
We first examine the resolution dependence of the observables in the Kerr-Newman limit. 
As shown in Fig. \ref{fig8}, while both observables converge at higher resolutions, the area observable $\tilde{A}_{\rm sh}$ exhibits significantly faster convergence and greater stability than $\tilde{\mathcal{R}}_{\rm ref}$.
Using the $1024\times1024$ results as the reference, the maximum relative differences are $0.348\%$ and $0.056\%$ for the area observable at $256\times256$ and $512\times512$, respectively, compared with $1.544\%$ and $0.477\%$ for the reference-circle radius.\footnote{For each observable and resolution, the maximum relative difference is the largest absolute difference from the corresponding $1024\times1024$ value, divided by that reference value, over all sampled charge points.}
Because $\tilde{\mathcal{R}}_{\rm ref}$ relies on a few local boundary pixels, it is susceptible to step-like fluctuations, whereas the global pixel integration of $\tilde{A}_{\rm sh}$ effectively smooths out such discretization errors. 
Based on this behavior, the shadow-observable curves in Fig. \ref{fig9} are computed at a grid resolution of $1024\times1024$ pixels.
The lower resolution used later for the large randomly sampled training ensemble is a separate computational choice and does not apply to Fig. \ref{fig9}.

We next quantify the contour deformation indicated in Fig. \ref{fig5} using the normalized axis observables.
Figure \ref{fig9} shows the variations of $\tilde d_{\max}$, $\tilde d_{\min}$, and $\tilde A_{\rm sh}$ with the Lorentz-violating parameter $\ell$ for different spin and charge configurations.
The horizontal separation $d_{(C,D)}$ remains essentially unchanged as $\ell$ varies, confirming that Lorentz violation does not modify the horizontal width of the shadow in the parameter range considered here, as already suggested by the auxiliary lines in Fig. \ref{fig5}.
The main deformation instead occurs in the vertical direction, characterized by $d_{(A,B)}$, which is stretched or compressed as $\ell$ varies.
Since $ d_{\max}$ and $ d_{\min}$ are defined as the larger and smaller values of $d_{(A,B)}$ and $d_{(C,D)}$, respectively, the apparent sharp kinks in the negative-spin curves arise when $d_{(A,B)}$ and $d_{(C,D)}$ exchange their roles as the longest and shortest axes.
These kinks therefore reflect a change in the axis classification rather than a discontinuous change of the shadow contour.
Moreover, the charged and uncharged curves are nearly parallel, indicating that the electric charge primarily changes the overall Kerr-Newman shadow scale, while no strong coupling between $Q$ and $\ell$ is observed in these axis observables.
The area observable $\tilde A_{\rm sh}$, as a global measure of the shadow region, provides an independent check of this picture.
The stretching and compression shown by the axis observables suggest that the signs of $\ell$ and of the branch-spin combination separate the rotating Lorentz-violating parameter space into three candidate morphology-response sectors,
\begin{equation}
\begin{aligned}
{\rm Sector\ I}:&\quad \ell>0,\quad a>0,\\
{\rm Sector\ II}:&\quad \ell<0,\quad a>0,\\
{\rm Sector\ III}:&\quad \ell<0,\quad a<0.
\end{aligned}
\label{eq:sectors}
\end{equation}
Equation \eqref{eq:sectors} specifies the representative domains sampled on the $(+)$ branch used in the image survey, rather than a branch-independent definition of the full physical sectors.
For $\ell>0$, no spin-sign dependence of the shadow morphology is resolved within the numerical resolution and parameter range considered here. We therefore sample the positive-spin domain as a representative of Sector I; this is a numerical choice and does not constitute an exact identification of the $a>0$ and $a<0$ configurations.
For $\ell<0$, by contrast, the two spin signs produce resolvably different stretched and compressed shadow families and are retained as separate sectors.
In a branch-independent notation, these two negative-$\ell$ sectors can be written as $\ell<0,sa>0$ and $\ell<0,sa<0$, where $s=\pm1$ denotes the branch sign discussed above.
To compare all three sectors on the same horizontal axis, we use the non-negative amplitude $\lambda=|\ell|$, so that $\ell=\lambda$ in Sector I and $\ell=-\lambda$ in Sectors II and III.
Accordingly, the Lorentz-violating response is evaluated with respect to $\lambda$, namely $\partial_\lambda=\partial_\ell$ in Sector I and $\partial_\lambda=-\partial_\ell$ in Sectors II and III.

This initially morphology-guided partition provides a convenient way to test whether the apparent near-superposition of the charge effect and the Lorentz-violating deformation persists across regions with different shadow morphologies.
At this stage, the three domains should be regarded as candidate morphology-response sectors: their relevance to inference is not inferred from visual morphology alone, but is tested below using observable sensitivities, mixed parameter responses, degeneracy structures, and sector-restricted posteriors.
Although Fig. \ref{fig9} already provides a useful comparison between selected charged and uncharged configurations, it only samples a few representative parameter choices and therefore cannot by itself establish whether $\tilde Q$ and $\ell$ are generally independent in the shadow observables.
We therefore perform a two-parameter scan in the $(\tilde Q,\lambda)$ plane at fixed spin and examine the possible mixed dependence between the electric charge and the Lorentz-violating deformation amplitude within the three sectors defined above.
Since the normalized area observable $\tilde A_{\rm sh}$ shows the best convergence behavior among the observables considered above, we use it as the representative diagnostic in this test.
Specifically, we compare the response of $\tilde A_{\rm sh}$ to $\lambda$ at different values of $\tilde Q$, and its response to $\tilde Q$ at different values of $\lambda$.
Equivalently, we examine whether the derivative curves $\partial \tilde A_{\rm sh}/\partial \lambda$ and $\partial \tilde A_{\rm sh}/\partial \tilde Q$, evaluated numerically from the parameter scan, depend on the value of the other parameter.
If these derivative curves remain unchanged when the other parameter is varied, $\tilde A_{\rm sh}$ can be regarded as approximately separable in $\tilde Q$ and $\lambda$ within a given sector; otherwise, their deviations indicate a mixed dependence between the two parameters.
To quantify the deviation between derivative curves, we take the derivative curve at $\tilde Q=0$ as the reference for the $\lambda$-direction scan and compare it with those obtained at finite charge.
For each value of $\tilde Q$, we define the root-mean-square deviation
\begin{equation}
\mathcal{C}_\lambda (\tilde{Q})={\sqrt{\tfrac{1}{N} \sum_i \big[\pp_\lambda \tilde{A}_{\rm sh}(\tilde{Q},\lambda_i) -\pp_\lambda \tilde{A}_{\rm sh}(0,\lambda_i)  \big]^2 }}.
\end{equation}
Here $N$ is the number of sampled points along the $\lambda$ direction.
Similarly, for the $\tilde Q$-direction scan, we take the derivative curve at $\lambda=0$ as the reference and define the root-mean-square deviation $\mathcal{C}_{\tilde Q}(\lambda)$ in the same way.
These quantities measure the absolute change in the response to one parameter when the other parameter is varied, and therefore provide a quantitative diagnostic of possible mixed $\tilde Q$-$\lambda$ dependence within each candidate morphology-response sector.

\begin{figure}[th]
\centering
\includegraphics[width=0.9\linewidth]{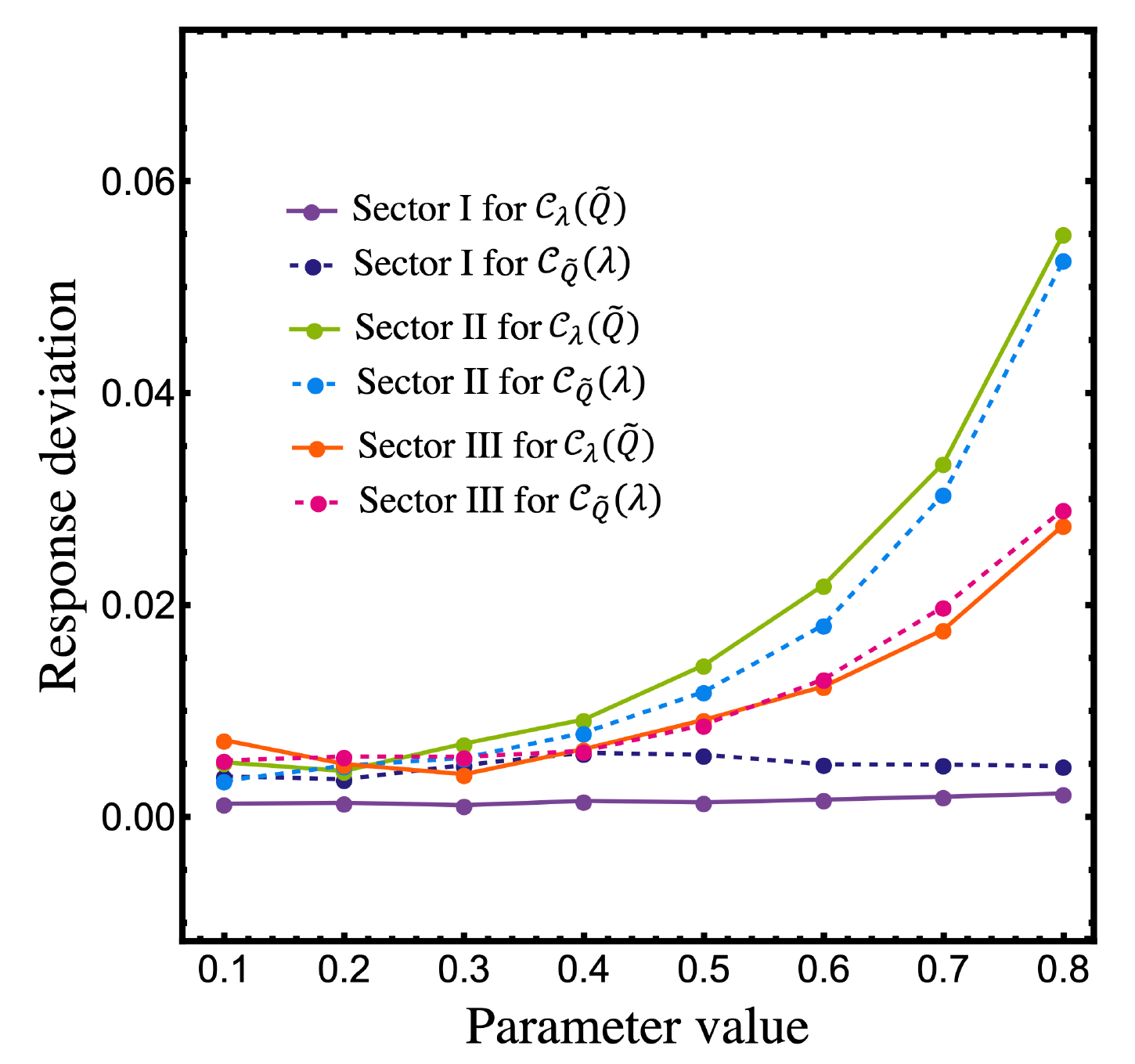}
\caption{Root-mean-square response deviations used to diagnose possible mixed $\tilde Q$-$\lambda$ dependence in the three candidate morphology-response sectors specified in Eq. \eqref{eq:sectors}. 
The spin magnitude is fixed at $|a/M|=0.67$, and $\lambda=|\ell|$. 
Solid and dashed curves denote $\mathcal{C}_\lambda(\tilde Q)$ and $\mathcal{C}_{\tilde Q}(\lambda)$, respectively, with the horizontal axis representing the corresponding argument.}
\label{fig10}
\end{figure}

As shown in Fig. \ref{fig10}, the mixed response exhibits a clear sector dependence.
In Sector I, both $\mathcal{C}_\lambda(\tilde Q)$ and $\mathcal{C}_{\tilde Q}(\lambda)$ remain at the level of the numerical fluctuation over the whole parameter range considered here. 
Within the resolution and scan range of the present ray-tracing calculation, Sector I can therefore be regarded as an effectively separable region for the area observable: the charge mainly changes the Kerr-Newman shadow scale, while the positive Lorentz-violating deformation contributes independently and produces no resolvable mixed response.
The situation changes qualitatively in the negative-$\ell$ sectors. 
For Sectors II and III, both response deviations grow well above the numerical fluctuation level as the control parameter increases. 
This means that the $\lambda$-response of the shadow area depends sensitively on $\tilde Q$, and conversely that the charge response depends sensitively on $\lambda$. 
Thus the charge and Lorentz-violating effects are no longer separable in these sectors, but show a pronounced mixed dependence in the shadow area observable.
This sector dependence can be understood from the metric structure. 
The Lorentz-violating radial-angular coupling is proportional to the branch-spin-Lorentz-violation combination, while the electric charge enters this coupling indirectly through the radial function in its denominator. 
Changing $\tilde Q$ therefore modifies the near-horizon radial structure that modulates the Lorentz-violating $r$-$\theta$ coupling, and hence affects the critical photon region and the projected capture boundary.
For positive $\ell$, this mixed contribution remains below the numerical sensitivity of the present calculation. 
For negative $\ell$, however, it becomes resolvable in the strong-field regime, especially in Sector II, where both diagnostics exhibit the largest deviations.
This confirms the need for the boundary-based observables introduced above: visually similar shadow images can correspond to different parameter responses once the sector structure is taken into account.
For the shadow-area observable and within the numerical range considered here, the response analysis therefore supports interpreting the initially morphology-guided domains as distinct morphology-response sectors. Their inference-level relevance is further assessed below through the sector-restricted posterior and degeneracy structures.




\section{Observation limits from EHT}\label{sec3}

The forward modeling in the previous section provides a map from the spacetime parameters and the viewing geometry to the shadow observables. 
In this section, we use this map in the inverse direction and constrain the charged Lorentz-violating black hole using the published EHT limits on the shadow observables.
We restrict our analysis to M87* because the available Sgr~A* measurements do not provide an axis-ratio constraint directly comparable to the one used here.
The inference parameters include the intrinsic spacetime parameters of the M87* black-hole template, such as $a$ and $Q$, the viewing parameter $\theta_0$, and the Lorentz-violation parameter $\ell$ of the gravitational theory.
The candidate morphology-response partition introduced in Eq. \eqref{eq:sectors} plays a central role in the inference. 
As shown by the forward-modeling analysis, Sector I behaves as an effectively separable region for the area observable, whereas Sectors II and III exhibit a resolvable mixed $\tilde Q$-$\lambda$ dependence. 
Although global Bayesian inference remains formally applicable, its marginalization can obscure sector-specific constraints; we therefore condition the posterior on each morphology-response sector.

The choice of observables should also reflect the sector structure. 
In Sector I, where the shadow contour remains close to the Kerr-Newman-like profile, both the reference-circle radius and the area-equivalent radius are meaningful scale estimators. 
However, in Sectors II and III, where droplet-like and elliptical shadows appear, the reference circle is no longer a reliable primary size measure.
We therefore use the area-equivalent radius as the baseline theoretical shadow-size proxy, while the reference-circle radius is kept in Sector I only as a consistency check.
Its angular-diameter conversion is compared with the EHT ring diameter, which is adopted here as the corresponding observational proxy.
Together with the fractional deviation $\delta^{\rm obs}$ from the Schwarzschild diameter scale and the axis ratio $\mathcal{D}_{\rm ax}^{\rm obs}$, this angular size proxy defines the observable vector
\begin{equation}\label{eq:eht_observable_vector}
\mathcal{O}^{\rm obs}
=
\big(\vartheta^{\rm obs},\delta^{\rm obs},
\mathcal{D}_{\rm ax}^{\rm obs}\big).
\end{equation}
These limits are taken from the EHT measurements of M87* \cite{EventHorizonTelescope:2019dse,EventHorizonTelescope:2019ggy,EventHorizonTelescope:2020qrl,EventHorizonTelescope:2021dqv,EventHorizonTelescope:2025vum}.
The M87* inputs are summarized in Table \ref{tab1}.
In the EHT analysis, the shadow size is reported as an angular observable, while the mass and distance enter as external calibration inputs in the conversion between angular and gravitational-radius scales \cite{EventHorizonTelescope:2019ggy}.
Following this treatment, the first three rows enter the likelihood, while $M_\bullet$ and $D_\bullet$ are fixed at their central values and used only to convert the dimensionless shadow size to an angular diameter.

\begin{table}[h]
\renewcommand{\arraystretch}{1.25}
\centering
\setlength\tabcolsep{1mm}{
\begin{tabular}{lccc}
\hline\hline
Quantity & Symbol  & \multirow{1}{*}{ Value(M87*) }&\multirow{1}{*}{ Ref.} 
\\
\hline
Angular diameter & $\vartheta^{\rm obs}$ & $(43.9\pm 0.6)~ \mu{\rm as}$ & 2025 \cite{EventHorizonTelescope:2025vum}  \\
Fractional deviation & $\delta^{\rm obs}$ & $-0.01\pm0.17$ & 2021 \cite{EventHorizonTelescope:2021dqv}  \\
Axis ratio & $\mathcal{D}_{\rm ax}^{\rm obs}$ & $<4:3$ & 2019 \cite{EventHorizonTelescope:2019dse} \\
Mass & $M_\bullet$ & $(6.5\pm0.7)\times 10^{9}\, M_\odot$ & 2019 \cite{EventHorizonTelescope:2019ggy} \\
Distance & $D_\bullet$ & $16.8^{+0.8}_{-0.7}\,{\rm Mpc}$ & 2019 \cite{EventHorizonTelescope:2019ggy} \\
\hline\hline
\end{tabular}}
\caption{EHT observational constraints and fixed calibration parameters for M87* used in the Bayesian inference.}
\label{tab1}
\end{table}

Adopting the general strategy of horizon-scale tests based on shadow observables \cite{Liu:2025wwq}, we use the parameter vector $\psi$ defined below Eq. \eqref{eq:normalized_observables} and denote its restriction to each morphology-response sector by
\begin{equation}
\psi_s\equiv \psi|_s,\qquad s={\rm I,II,III},
\end{equation}
where the subscript indicates that the prior range is restricted to the corresponding morphology-response sector.
The sector-restricted priors used in the MCMC analysis are summarized in Table \ref{tab2}.
All parameters are sampled with uniform priors within the listed intervals, together with the Kerr-Newman-type horizon condition $\tilde a^2+\tilde Q^2<1$.
Here $\lambda=|\ell|$ is used as the non-negative Lorentz-violation amplitude, with $\lambda_{\max}=0.99$.
For Sector I, no spin-sign dependence of the area observable is resolved within the scanned resolution and parameter range, so the positive-spin domain is sampled as a numerical representative. This sampling choice does not assume an exact equivalence between the two spin signs.
The inclination is restricted to $\theta_0\in[0,\pi/2]$ by the reflection symmetry of the shadow with respect to the equatorial plane.

\begin{table}[h]
\renewcommand{\arraystretch}{1.25}
\centering
\setlength\tabcolsep{2.5mm}{
\begin{tabular}{ccccc}
\hline\hline
\multirow{1}{*}{Sector} & $\tilde{a}$  & $\tilde{Q}$ & $\ell$ & $\theta_0$
\\
\hline
I & $[0,0.99]$ & $[0,0.99]$ & $[0,\lambda_{\max}]$ & $[0,\pi/2]$ \\
II & $[0,0.99]$ & $[0,0.99]$ & $[-\lambda_{\max},0]$ & $[0,\pi/2]$ \\
III & $[-0.99,0]$ & $[0,0.99]$ & $[-\lambda_{\max},0]$ & $[0,\pi/2]$ \\
\hline\hline
\end{tabular}}
\caption{Sector-restricted prior ranges used in the MCMC analysis.
All priors are uniform within the listed intervals.}
\label{tab2}
\end{table}

For each parameter point and each morphology-response sector, we first define a Schwarzschild-calibrated shadow-size factor from the shadow area,
\begin{equation}
\Xi_s(\psi_s)=
\sqrt{A_{\rm sh}^{\rm th}(\psi_s)/A_{\rm sh}^{\rm th}(\psi_{\rm Sch})},
\qquad s={\rm I,II,III}.
\end{equation}
This follows the Schwarzschild-calibration logic of Appendix A in Ref. \cite{Liu:2025wwq}: the Schwarzschild image supplies the reference scale, and the target shadow is compared with it through a dimensionless ratio.
For Sector I, we also compute the reference-circle analogue
\begin{equation}
\Xi_{{\rm ref},{\rm I}}(\psi_{\rm I})=
\mathcal{R}_{\rm ref}^{\rm th}(\psi_{\rm I})/
\mathcal{R}_{\rm ref}^{\rm th}(\psi_{\rm Sch}),
\end{equation}
which is used only to test the robustness of the Sector-I constraints against the choice of size estimator.
It is useful to write this scale factor as an effective dimensionless shadow radius,
\begin{equation}
R_{{\rm sh},s}^{\rm th}(\psi_s)=3\sqrt{3}\,\Xi_s(\psi_s),
\end{equation}
where $R_{{\rm sh},s}^{\rm th}$ is measured in units of the black-hole mass.
Because this quantity is a radius scale whereas the EHT size measurement in Table \ref{tab1} is quoted as an angular diameter, the comparison with $\vartheta^{\rm obs}$ requires an additional factor of two.
For M87* with mass $M_\bullet$ and distance $D_\bullet$, the theoretical angular diameter is therefore \cite{Amarilla:2011fx}
\begin{equation}
\begin{aligned}
\vartheta_s^{\rm th}(\psi_s)
\!&=2R_{{\rm sh},s}^{\rm th}(\psi_s)\frac{M_\bullet}{D_\bullet}  \\
\!&=1.974196\times10^{-5}R_{{\rm sh},s}^{\rm th}(\psi_s)
\left(\frac{M_\bullet}{M_\odot}\right)
\left(\frac{1{\rm kpc}}{D_\bullet}\right)\mu{\rm as}.
\end{aligned}
\end{equation}
Equivalently, if $\vartheta_{\rm Sch}=6\sqrt{3}M_\bullet/D_\bullet$ denotes the Schwarzschild angular diameter for M87*, one has $\vartheta_s^{\rm th}=\vartheta_{\rm Sch}\Xi_s$.
The corresponding fractional deviation from the Schwarzschild diameter scale is
\begin{equation}
\delta_s^{\rm th}(\psi_s)=\frac{R_{{\rm sh},s}^{\rm th}(\psi_s)}{3\sqrt{3}}-1
=\Xi_s(\psi_s)-1 .
\end{equation}
The ray-traced shadow factor $\Xi_s$ and the axis ratio are determined by the black-hole parameters and the viewing angle.

Corresponding to the data vector in Eq. \eqref{eq:eht_observable_vector}, the forward model gives the theoretical prediction vector
\begin{equation}\label{eq:theory_observable_vector}
\mathcal{O}_{s}^{\rm th}(\psi_s)=\big(\vartheta_s^{\rm th},\delta_s^{\rm th},
\mathcal{D}_{{\rm ax},s}^{\rm th}\big)(\psi_s),
\end{equation}
whose components are ordered in the same way as the EHT observables in Eq. \eqref{eq:eht_observable_vector}.
As a working approximation, we treat the three phenomenological EHT constraints as statistically independent and represent them using effective Gaussian likelihood terms, neglecting possible covariance among them.
The likelihood is then written as
\begin{equation}
\mathcal{L}_s(\mathcal{O}^{\rm obs}|\psi_s)
\propto
\exp\left[
-\frac{1}{2}\sum_i
\frac{\left(\mathcal{O}_{s,i}^{\rm th}(\psi_s)-\mathcal{O}_i^{\rm obs}\right)^2}{\sigma_i^2}
\right],
\end{equation}
where the index $i$ runs over the three EHT constraints listed in $\mathcal{O}^{\rm obs}$.
The posterior distribution in each sector is then obtained from Bayes' theorem,
\begin{equation}
P(\psi_s|\mathcal{O}^{\rm obs})
=\frac{\mathcal{L}_s(\mathcal{O}^{\rm obs}|\psi_s)\Pi_s(\psi_s)}
{\mathcal{Z}_s},
\end{equation}
where $\Pi_s(\psi_s)$ is the sector-restricted prior and $\mathcal{Z}_s$ is the Bayesian evidence.

In practice, direct ray tracing at every MCMC sample would be computationally expensive.
We therefore precompute the shadow observables on a large ensemble of ray-traced images in each sector and interpolate the resulting forward map during the MCMC sampling.
For each sector, we generate $4\times10^4$ shadow images from randomly sampled parameter points within the validated parameter ranges, using a screen resolution of $512\times512$ pixels.
Because the randomly sampled training set is substantially larger than the deterministic scan used for Fig. \ref{fig9}, we use $512\times512$ pixels for this ensemble. 
Figure \ref{fig8} supports the stability of the area-based size proxy at this resolution; residual discretization error is not propagated separately in the present likelihood.
The sectors are sampled independently, yielding conditional posteriors that retain response and degeneracy structures potentially blurred by global marginalization.
For presentation only, the Sector-III spin posterior is plotted in terms of the positive quantity $|\tilde a|$, while the sampled parameter itself remains negative as specified in Table \ref{tab2}.
Having constructed this sector-dependent forward map, we now present the MCMC constraints obtained from the EHT observables listed in Table \ref{tab1}.

\begin{figure}[th]
\centering
\includegraphics[width=1\linewidth]{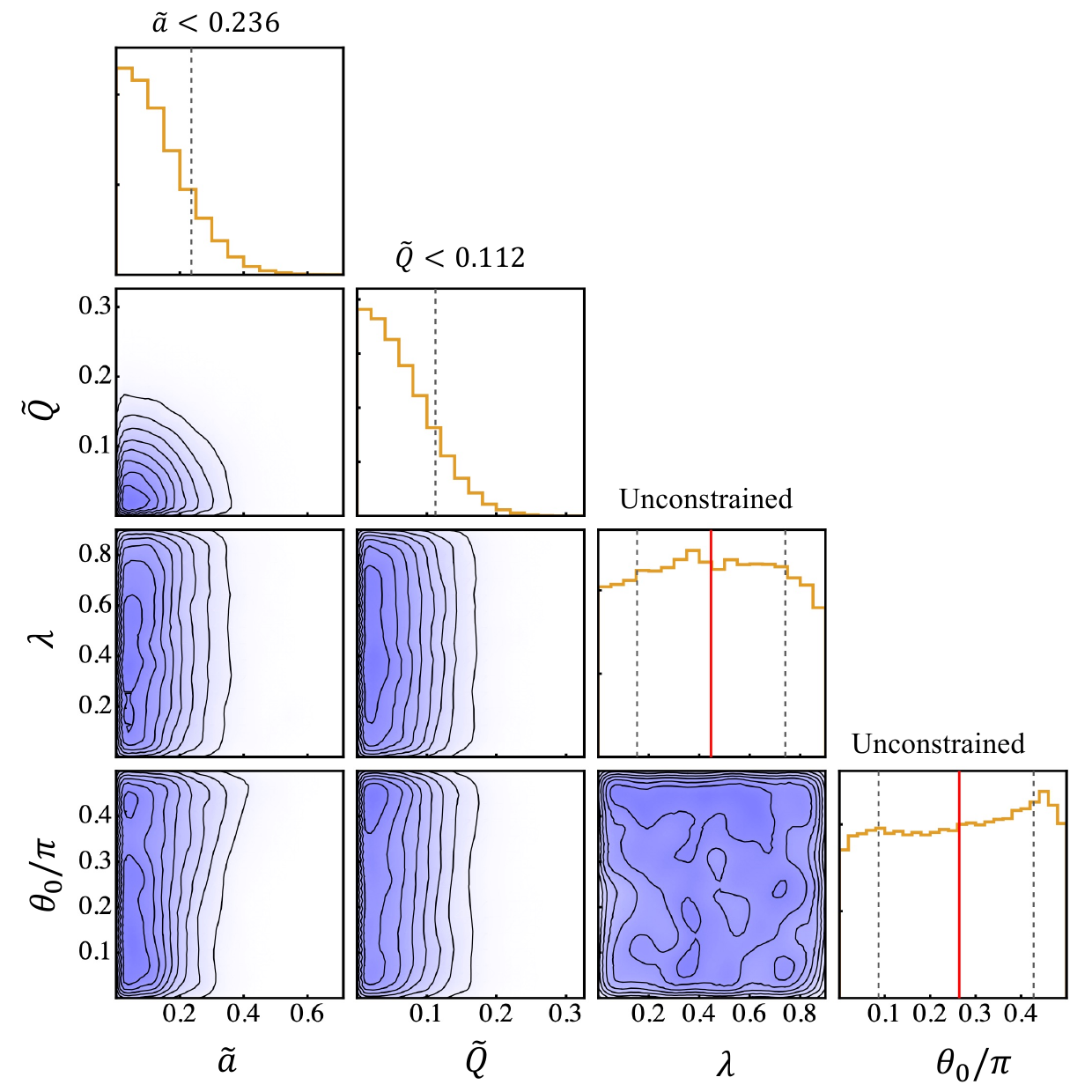}
\caption{Corner plot showing the one- and two-dimensional posterior probability distributions for Sector I, constrained by the M87* observables in Table \ref{tab1}.
The dashed lines indicate the 68\% credible limits, while the red solid lines mark the posterior medians when well defined.}
\label{fig11}
\end{figure}

Figure \ref{fig11} shows the posterior constraints in Sector I. 
In this sector, the spin and electric charge are constrained mainly through upper limits, with $\tilde a<0.236$ and $\tilde Q<0.112$ at the 68\% credible level, whereas the parameter $\ell$ and the viewing angle $\theta_0$ remain effectively unconstrained. 
This behavior is consistent with the forward-modeling results in Sec. \ref{subsec:observables}. 
The EHT angular-size proxy provides the dominant constraint on the shadow scale, and within the sampled Kerr-Newman-like Sector-I templates both spin and charge tend to reduce the area-equivalent shadow size.
Their combined effect therefore translates into marginalized upper bounds on $\tilde a$ and $\tilde Q$.
By contrast, the positive Lorentz-violating parameter produces only a mild variation in the area observable and does not generate a clearly resolvable non-circular deformation of the shadow boundary within the present ray-tracing resolution and parameter range.
As a result, the present EHT constraints cannot distinguish different values of $\ell$ or provide an independent constraint on the viewing angle.
Compared with fixed-parameter scans or contour-based estimates \cite{Vagnozzi:2019apd,Bambi:2019tjh,Long:2019nox}, the present analysis varies the spin, charge, Lorentz-violation parameter, and viewing angle simultaneously and propagates the observational uncertainties through the likelihood.
The upper bounds on $\tilde a$ and $\tilde Q$ in Fig. \ref{fig11} should therefore be understood as marginalized Bayesian constraints within the Sector-I morphology-response sector, while the absence of a meaningful constraint on $\ell$ reflects the weak sensitivity of the Sector-I shadow observables to positive Lorentz violation.

\begin{figure}[th]
\centering
\includegraphics[width=1\linewidth]{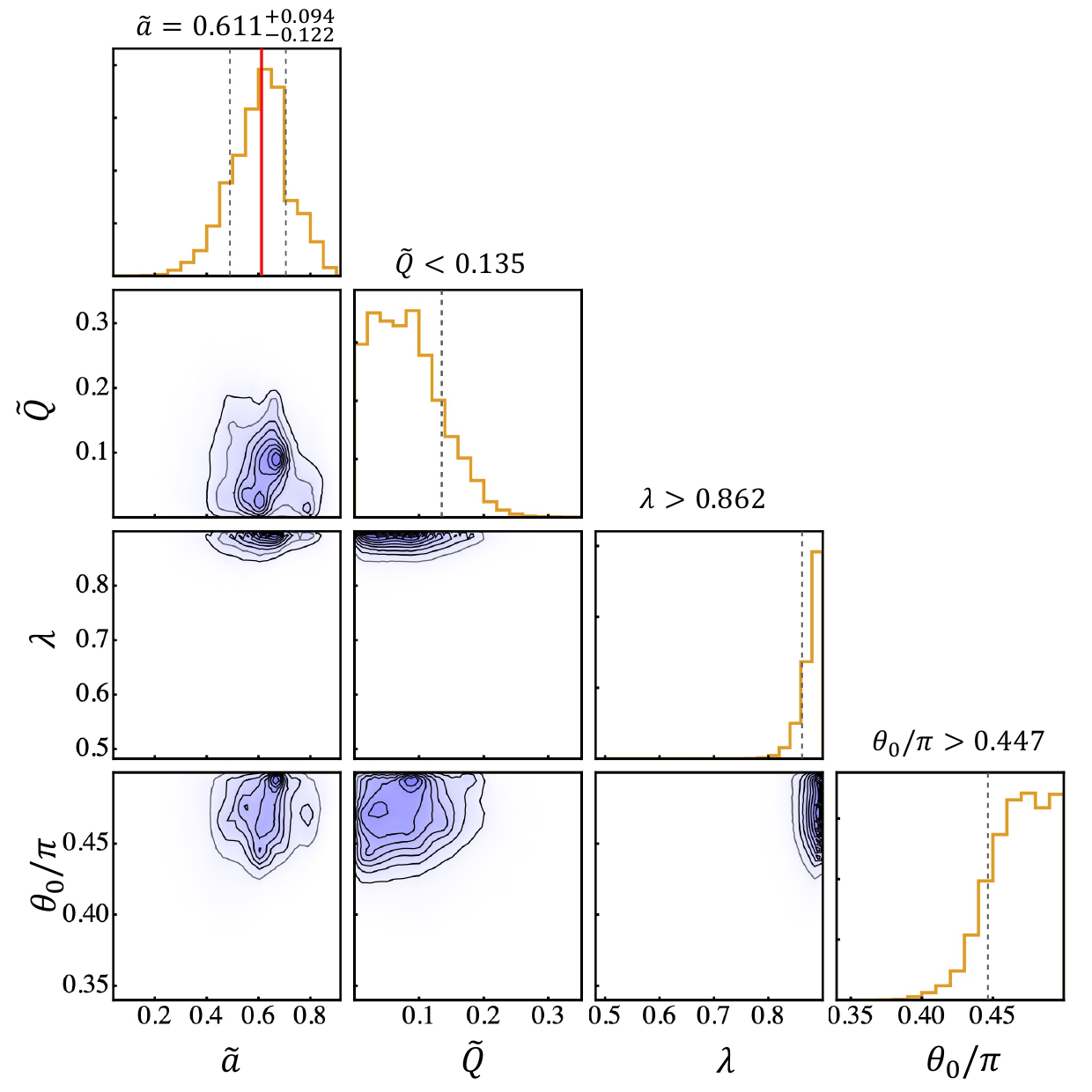}
\caption{Corner plot showing the one- and two-dimensional posterior probability distributions for Sector II, constrained by the M87* observables in Table \ref{tab1}.
The dashed lines indicate the 68\% credible limits, while the red solid lines mark the posterior medians when well defined.}
\label{fig12}
\end{figure}

The negative-$\ell$ sectors show qualitatively different inference behavior.
As shown in Fig. \ref{fig12}, Sector II gives a localized posterior for the spin, $\tilde a=0.611^{+0.094}_{-0.122}$, while the electric charge is still mainly constrained through an upper bound, $\tilde Q<0.135$.
The parameter $\ell$ is driven toward the large negative region, corresponding to $\lambda>0.862$, or equivalently $\ell<-0.862$, at the 68\% credible level.
This is consistent with the forward-modeling result that the negative-$\ell$ sectors are much more sensitive to Lorentz violation than Sector I.
On the representative $(+)$ branch, Sector II is the vertically stretched, more elliptical negative-$\ell$ family and exhibits the largest resolved mixed $\tilde Q$-$\lambda$ response in Fig. \ref{fig10}. Unlike Sector I, its posterior has a localized spin, a lower-bounded $\lambda$, an upper-bounded $\tilde Q$, and sector-specific joint contours. Without a likelihood-ablation analysis, we do not assign these features to any single EHT observable.
\scopeedit{The posterior in Fig. \ref{fig12} gives the inclination preferred under the Sector-II morphology-response hypothesis.}

\begin{figure}[th]
\centering
\includegraphics[width=1\linewidth]{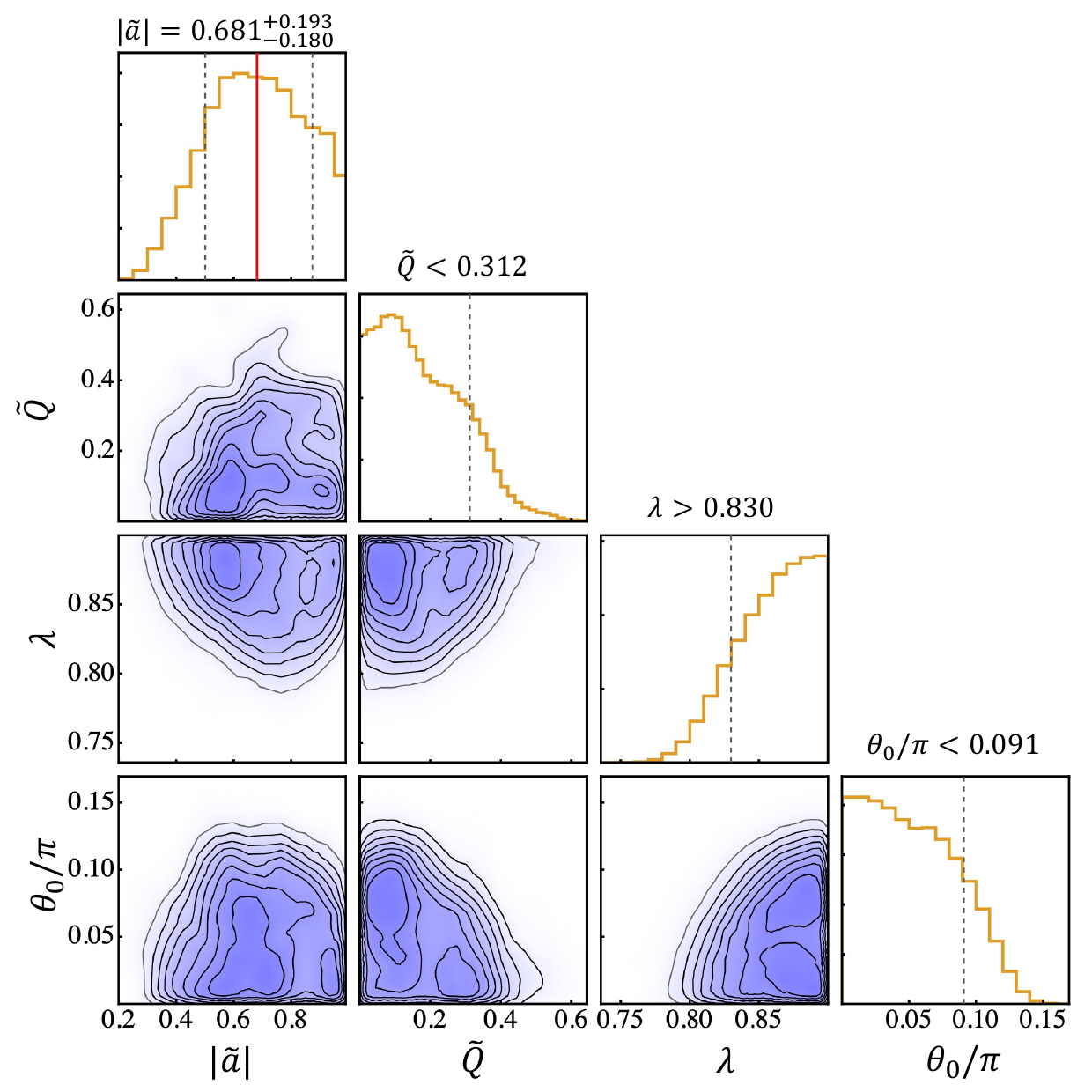}
\caption{Corner plot showing the one- and two-dimensional posterior probability distributions for Sector III, constrained by the M87* observables in Table \ref{tab1}; the spin axis is shown as $|\tilde a|$.
The dashed lines indicate the 68\% credible limits, while the red solid lines mark the posterior medians when well defined.}
\label{fig13}
\end{figure}

Figure \ref{fig13} shows the corresponding result for Sector III.
Since this sector has negative spin, the plotted spin variable is $|\tilde a|$, although the sampled spin parameter itself is negative.
The posterior favors $|\tilde a|=0.681^{+0.193}_{-0.180}$, with $\tilde Q<0.312$ and $\lambda>0.830$, or equivalently $\ell<-0.830$, at the 68\% credible level.
Sector III is the vertically compressed, droplet-like negative-$\ell$ family on the representative $(+)$ branch. It retains a mixed $\tilde Q$-$\lambda$ response, but its joint contours and inclination distribution differ from Sector II; the two negative-$\ell$ sectors therefore do not share a common posterior structure.
Compared with Sector II, Sector III yields a substantially different inclination posterior.
This does not mean that the physical source inclination is free to change between sectors.
Rather, Sectors II and III represent two different morphology-response hypotheses for the same source and define different mappings from the intrinsic parameters and viewing geometry to the projected shadow observables.
Their different inclination posteriors are therefore consistent with the fixed-branch spin-reversal asymmetry of the forward-modeled shadows. However, because the present analysis does not perform a likelihood-ablation or parameter-response decomposition for the inclination, we do not attribute the posterior difference uniquely to this asymmetry.
A possible qualitative connection with the macroscopic optical nonreciprocity discussed in Ref. \cite{Liu:2026optical} may merit further investigation.
Across all sectors, the charge parameter remains primarily bounded from above rather than being positively selected, whereas the Lorentz-violating parameter becomes strongly constrained only in the negative-$\ell$ morphology-response sectors.

Taken together, Figs. \ref{fig11}--\ref{fig13} show that the EHT constraints are strongly sector dependent.
Sector I mainly provides upper bounds on the Kerr-Newman-like scale-reducing parameters $\tilde a$ and $\tilde Q$, while leaving $\ell$ and $\theta_0$ weakly constrained.
In the negative-$\ell$ sectors, the sector-restricted posteriors are concentrated toward large $|\ell|$, in contrast to the weakly constrained positive-$\ell$ parameter in Sector I. This difference is consistent with the stronger observable response to Lorentz violation found in the forward-modeling analysis.
Moreover, Sectors II and III yield different inferred viewing inclinations, a result that is consistent with, but not uniquely attributable to, the fixed-branch spin-reversal asymmetry of the forward-modeled shadows.
Together with the nonuniform forward responses, these posterior structures support separate sector treatment. A global fit would mix the upper-limit behavior of Sector I with the localized spin, lower-bounded $|\ell|$, and different inclination structures of Sectors II and III.

\par\vspace{0.8\baselineskip}
\section{Conclusion and discussion}

In this work, we investigated black-hole shadow phenomena generated by an exact Kerr-Newman-like rotating geometry with a radial-angular cross term.
Although the Lorentz-violating theory provides a concrete realization of this geometry, the essential ingredient is the off-diagonal component $g_{r\theta}$, which couples the radial and angular photon motion and makes the shadow morphology sensitive to the relative orientation encoded by the branch sign and spin direction, as well as to the Lorentz-violating parameter. 
The resulting dynamics is therefore not a simple Kerr-Newman deformation governed only by the usual radial function. 
Our backward ray-tracing analysis shows that, for positive $\ell$, the shadow boundary remains close to the Kerr-Newman-like contour in the parameter range considered here, although the surrounding lensing pattern is modified. For negative $\ell$, however, the coupling can generate qualitatively different profiles, including droplet-like and elliptical shadows. Since these deformations depend on the relative sign between the spin and the cross term, spin reversal within a fixed branch does not merely reflect the image but can change the projected morphology itself. This spin-reversal asymmetry constitutes a characteristic optical signature of the radial-angular coupling.

Although the shadow contours evolve continuously across parameter space, their morphological responses and parameter-degeneracy structures are not uniform. Global inference remains formally possible, but its marginalized posterior may obscure these sector-specific structures. We introduced boundary-based observables, including the area-calibrated shadow size, the longest and shortest axes, and the axis ratio, and used the resulting morphology and response structures to identify three morphology-response sectors. The area observable is more stable than the reference-circle radius for irregular contours and makes the inference applicable to strongly noncircular profiles without forcing them into a common reference-circle template. The mixed-response analysis further shows that, for the shadow-area observable and within the numerical range considered here, the charge and Lorentz-violating responses are effectively separable in Sector I but exhibit a resolvable mixed dependence in the negative-$\ell$ sectors. Motivated by this structure, we conditioned the Bayesian analysis on the three morphology-response sectors and reported the resulting sector-specific posteriors. The distinct posterior and degeneracy structures provide additional inference-level support for this partition and show that the inferred parameter correlations depend on the underlying shadow morphology and response regime.

More broadly, our results show that a continuously varying family of black-hole shadows can exhibit qualitatively distinct response and inference regimes. This finding motivates morphology-resolved, rather than globally uniform, parameter estimation for non-Kerr spacetimes with complex shadow structures. The sequence from cross-coupled geometry to modified photon dynamics, nonstandard shadow morphology, and sector-resolved inference provides a useful strategy for studying similar effects in other beyond-Kerr spacetimes. Future high-resolution black-hole imaging, together with more realistic emission and image-reconstruction models, will be essential for determining whether such morphology-dependent templates can be distinguished observationally from Kerr-Newman and other modified-gravity black holes.

\acknowledgments
The authors thanks Dr. Yang Liu for providing important discussions on the paper.
This work is supported in part by the National Natural Science Foundation of China (Grants No. 12605108, 12547147, 12475056 and 12247101), the Fundamental Research Funds for the Central Universities under Grant Nos.~lzujbky-2024-jdzx06 and lzujbky-2025-jdzx07, Gansu Province’s Top Leading Talent Support Plan, the Natural Science Foundation of Gansu Province (Nos.~22JR5RA389 and 25JRRA799), the ``111 Center'' (Grant No.~B20063), and the China Postdoctoral Science Foundation (Grant No. 2025M783393).

\appendix

%

\end{document}